\documentclass{jfm}
\usepackage{graphicx}
\usepackage{epstopdf,epsfig}
\usepackage{newtxtext}
\usepackage{newtxmath}
\usepackage{natbib}
\usepackage{hyperref}
\usepackage{subcaption}
\usepackage[normalem]{ulem}
\hypersetup{
    colorlinks = true,
    urlcolor   = blue,
    citecolor  = black,
}

\usepackage{xcolor}
\definecolor{myblue}{HTML}{2E59A7} 
\definecolor{myred}{HTML}{D23918} 
\definecolor{myblack}{HTML}{151D29} 
\definecolor{myyellow}{HTML}{E5A84B} 
\definecolor{mygreen}{HTML}{057748} 
\definecolor{mybrown}{HTML}{9F6027} 
\definecolor{mypink}{HTML}{FF0097} 
\newcommand{\cgreen}[1]{{\color{mygreen} {#1}}}

\newcommand{\cred}[1]{{\color{red} {#1}}}

\newcommand{\RomanNumeralCaps}[1]

\begin{document}

\title{Opposition control applied to turbulent {wings}}

\author{Yuning Wang\aff{1}, 
Marco Atzori\aff{2}, 
\and Ricardo Vinuesa\aff{1}\corresp{\email{rvinuesa@mech.kth.se}}}

\affiliation{\aff{1}FLOW, Engineering Mechanics, KTH Royal institute of Technology, 100 44 Stockholm, Sweden
\aff{2}Department of Aerospace Science and Technologies (DAER), Politecnico di Milano, Via La Masa 34, 20156 Milan, Italy}

\maketitle

\begin{abstract}
We carried out high-resolution large-eddy simulations (LESs) to investigate the effects of opposition control (OC) 
on turbulent boundary layers (TBLs) over a wing at a chord-based Reynolds number (${Re}_c$) of $200 \ 000$. 
Two cases were considered: flow over the suction sides of the NACA0012 wing section at an angle of attack of $0^{\circ}$, 
and the NACA4412 wing section at an angle of attack of $5^{\circ}$. 
These cases represent TBLs subjected to mild and strong nonuniform adverse pressure gradients (APG), respectively. 
\nobreak
First, we assessed the control effects on the streamwise development of TBLs and the achieved drag reduction. 
Our findings indicate that the performance of OC in terms of friction-drag reduction significantly diminishes as the APG intensifies. 
Detailed analysis of turbulence statistics reveals that this is directly connected to the intensified wall-normal convection caused by {the} strong APG. 
Opposition control, {which is designed} to mitigate {the near-wall fluctuations}, 
results in {the} attenuation of {the outer-scaled} outer peak of streamwise velocity fluctuations and the production term of the turbulent kinetic energy budget.
The formation of the so-called virtual wall~\citep{hammond_observed_1998} is also confirmed in this study, 
{where the balance between viscous diffusion and dissipation at the virtual wall plane is remarkably similar to that at the physical wall.} 
Additionally, spectral {analyses} indicate that the wall-normal transport of 
small-scale structures to the outer region due to {the} APG has a {detrimental} impact on the performance of OC.
{Uniform blowing and body-force damping were also examined 
to understand the differences between {the various} control schemes.
Despite the distinct performance of friction-drag reduction, 
the effects of uniform blowing are akin to those induced by a stronger APG, 
while the effects of body-force damping exhibit similarities to those of OC 
in terms of the streamwise development of {the TBL} {although there are differences in the turbulent statistics}.}
\nobreak
To authors' best knowledge, the present study stands for the first in-depth analysis of the effects of OC 
applied to TBL subjected to nonuniform {APGs} with complex geometries.
\end{abstract}

\section{Introduction}
\label{sec:introduction}
{Turbulent-flow} control on aircraft wings for drag reduction is of significant importance in aviation as it enhances the aerodynamic performance of aircraft, 
contributing to emission control and yielding substantial economic and environmental benefits~\citep{fukagata_reviewCTRL_2024,wang_reviewCTRL_2024}. 
The primary source of drag is viscous drag~\citep{Anderson_fundamental_1991}, 
which results from the interactions of the turbulent boundary layer (TBL) on the wing surface, 
accounting for approximately $50\%$ of the total drag~\citep{abbas_TBLCTRLReview_2017}. 
Therefore, controlling viscous drag is crucial for reducing fuel consumption in commercial aircraft. 
Additionally, TBLs over airfoil surfaces often experience adverse or favorable pressure gradients (APG/FPG), 
adding complexity to the turbulent flow. 
This underscores the challenge and the urgent need for effective {drag-reduction} strategies on wings.

Over the past decades, numerous {turbulent-flow} control methods have been proposed. 
These methods can be categorized as passive or active control, 
depending on whether they require an additional energy source~\citep{gad_modern_1996}. 
Passive methods, such as using riblets, 
modify wing configurations without additional energy to reduce skin-friction drag~\citep{viswanath_riblets_2002}. 
Active flow control (AFC), on the other hand, 
manipulates the flow field using additional energy~\citep{vinuesa_flow_2022}.

AFC can be performed in a predetermined open-loop manner where the control input is independent {of} the flow state~\citep{brunton_closed-loop_2015}. 
One of the celebrated open-loop methods is uniform blowing/suction, which {was} firstly proved by~\citet{hwang_proofuniformblowing_1997}.
High-fidelity numerical simulations conducted by~\citet{vinuesa_skin_2017} and subsequent studies~\citep{atzori_aerodynamic_2020,atzori_uniform_2021} 
have extensively investigated turbulent flow around a NACA4412 wing section at Reynolds numbers up to $400 \ 000$, 
exploring various combinations of blowing and suction. 
These studies reported a maximum reduction of $8\%$ in friction drag and $5\%$ in total drag~\citep{atzori_aerodynamic_2020}. 
They concluded that uniform blowing reduces {skin-friction} drag while increasing turbulent fluctuations, 
whereas uniform suction has the opposite effect. 
This has been confirmed experimentally~\citep{eto_uniformClarkY_2019} 
and in other numerical studies on {zero-pressure-gradient} (ZPG) TBLs~\citep{fukagata_lower_2009,stroh_influence_2012}.
Additionally,~\citet{albers_transsurfwave_2019} carried out LES for assessing the effect of transversal surface waves on an airfoil at ${Re}_c = 400 \ 000$, 
where a drag reduction of $7.5\%$ {was} achieved.

In contrast, AFC can also be reactive, implemented in a closed-loop manner, 
where the control action is based on real-time {flow-state} sensing~\citep[as in \textit{e.g.}][]{vinuesa_flow_2022}. 
{Here, machine-learning methods have been used for control and sensing~\citep{yousif_ML_GAN_2023}, leading to very promising results.} 
Reactive schemes generally exhibit higher efficiency than open-loop methods~\citep{stroh_comparison_2015}, 
as demonstrated in experimental studies~\citep{king_active_2007,king_active_2010}. 
One promising reactive approach is {opposition control~\citep[OC]{choi_active_1994}}, 
which reduces friction drag by using blowing and suction to suppress streamwise vortices, 
interrupting the self-sustaining processes near the wall, such as sweep and ejection events~\citep{kravchenko_relation_1993,orlandi_generation_1994}. 
The imposed wall velocity, as {the control name} indicates, opposes the detected velocity fluctuations at a prescribed sensing plane $y_{s}$, 
{and it is setup to ensure} a zero-net-mass-flux condition at the wall.

OC was first proposed by~\citet{choi_active_1994} through direct numerical simulations (DNS) of fully developed turbulent channel flow (TCF) 
at ${Re}_{c} \approx 3300$ based on the centerline velocity $U_c$ and channel half-width $\delta$. 
In that work, various implementations of OC, including \textit{v}-control, \textit{w}-control and combined-control methods, 
{were} explored. {The same strategies were subsequently} revisited by~\citet{wang_active_2016}. 
Further studies {identified} the optimal locations for sensing planes and control amplitude adjustments 
to achieve higher drag reduction~\citep{hammond_observed_1998,chung_effectivenessTCF_2011}, 
where the control mechanism is outlined as the formulation {of the} so-called virtual wall~\citep{hammond_observed_1998}.
Additionally, \citet{ge_dynamicTCF_2017} investigated the effect of OC on redistributing {energy across the terms of the} {turbulent-kinetic-energy} (TKE) {budget}. 
Interestingly, a recent study has also utilized deep learning for blowing-only OC, 
demonstrating the potential for advanced control schemes~\citep{li_blowing-only_2021}. 
Experimental investigations by~\citet{stroh_influence_2012} focus on the limitations {in the} practical implementation of OC.

Spatially developing TBLs under OC exhibit different responses compared to TCF due to stronger fluctuations in spanwise and wall-normal velocities, 
as well as pressure~\citep{jimenez_TBLvsTCF_2010}. 
Numerical studies using LES have shown significant drag reduction and the formation of a virtual wall~\citep{pamies_response_2007}. 
Systematic comparisons between TBL and TCF under OC have highlighted 
the importance of {control-region} determination and the challenges 
posed by stronger near-wall fluctuations in TBLs~\citep{stroh_comparison_2015,xia_direct_2015}. 
Practical implementations, such as wall deformation combined with OC, 
have also been proposed and investigated~\citep{pamies_opposition_2011,dacome_oppositionTBL_2024}.

Despite these advances, most studies on TBL under OC have focused on {ZPG} conditions, 
which do not reflect the realistic scenarios on aircraft wings where nonuniform pressure gradients and finite aerodynamic bodies are involved. 
This {fact} motivates the present study, which aims to explore the interaction between OC and nonuniform APG 
by analyzing the properties of developing TBLs over the suction sides of airfoils with different geometries at moderate Reynolds numbers. 
High-resolution LES was performed on a NACA0012 wing section {at an} angle of attack ($AoA$) of $0^{\circ}$ 
and a NACA4412 wing section at $AoA = 5^{\circ}$, both at Reynolds number of ${Re}_c = 200 \ 000$ 
{($Re_c = U_{\infty} c/\nu$, where $c$ is the chord length and $\nu$ is the kinematic viscosity)}, 
representing mild and strong APG intensities, respectively~\citep{tanarro_effect_2020}. 
In the present study, we compare OC and uniform blowing as {examples of} predetermined and reactive AFC schemes for drag reduction,  
and {include} body-force damping results on the suction side of {a} NACA4412~\citep{atzori_uniform_2021} 
to enhance the understanding of reactive control mechanisms applied {to} {TBLs} subjected to {nonuniform} {APGs}.
\nobreak
To the best of the authors' knowledge, this paper represents the first numerical study {documenting the application of} opposition control 
{to this type of flow}.

{The} paper is organized as follows. 
In $\S$~\ref{sec:method}, we introduce the numerical setup and the implementation of control schemes. 
In $\S$~\ref{sec:bl_develop}, we assess the control effects on streamwise development of the boundary 
layer as well as the achieved drag reduction in terms of skin-friction drag and total drag.
Then in $\S$~\ref{sec:stat}, we present the results of 
the wall-normal profiles of turbulence statistics including 
mean velocity and fluctuations as well as the budgets of the turbulent kinetic energy.
In $\S$~\ref{sec:spectra}, we demonstrate the results of one- and two-dimensional spectral analysis. 
Finally, in $\S$~\ref{sec:conclusions} {the conclusions of this work are presented}.

\section{Methodology}\label{sec:method}
\subsection{High-resolution large-eddy simulation (LES)}\label{sec:simulation}
The numerical investigation of control schemes is performed using high-resolution LES of TBL on 
the wing section NACA0012 at $0^{\circ}$ angle of attack ($AoA$) and the wing section NACA4412 with $AoA$ of $5^{\circ}$, 
both at a Reynolds number of ${Re}_c = 200 \ 000$. 
{As described below, the present high-resolution LES employs a grid very close to that of a DNS.}

{All simulations are carried out with} the incompressible {Navier--Stokes} solver Nek5000~\citep{nek5000}, {which is} 
based on the spectral-element method (SEM). 
The SEM ensures the accuracy of the solution while maintaining computational efficiency~\citep{demoura_semadvantage_2024}. 
Specifically, the computational domain is discretized into hexahedral elements where velocity and pressure are represented by Lagrange interpolants. 
Following the $\mathbb{P}_N \mathbb{P}_{N-2}$ formulation~\citep{maday_spectral_1989}, 
if the polynomial order is $N-1$, 
the velocity is defined on $N^3$ points per element distributed according to the {Gauss--Lobatto--Legendre} (GLL) quadrature rule, 
while the pressure is defined on a staggered grid of ${(N-2)}^3$ points with the {Gauss--Legendre} (GL) distribution. 
For time advancement in the incompressible {Navier--Stokes} equations, 
the nonlinear terms are solved by explicit third-order extrapolation (EXT3), 
whereas the viscous terms are solved using an implicit third-order backward scheme (BDF3). 
To address aliasing errors, overintegration is implemented by oversampling the nonlinear terms 
by a factor of $3/2$ of the adopted polynomial order in each direction.
This simulation framework has been previously used for high-fidelity simulations of complex turbulent flows, 
see {\it e.g.} the work by \citet{noorani_aspect_2016}.

The high-resolution LES captures the {effects of the} smallest turbulent scales via a subgrid-scale (SGS) model 
based on a time-independent relaxation-term filter {developed by~\citet{negi_unsteady_2018}.} 
The filtering operation is performed implicitly through a volume force accounting for 
the dissipation of unresolved turbulent scales, 
{and this operation is conducted ensuring that continuity is preserved.} 
Including the LES filter, the governing equations of the system are the 
incompressible continuity and momentum equations in nondimensional form, expressed as:
\begin{align}
    \frac{\partial u_i}{\partial x_i} &= 0, \\
    \frac{\partial u_i}{\partial t} + u_j \frac{\partial u_i}{\partial x_j} &= 
    -\frac{\partial p}{\partial x_i} + \frac{1}{{Re}_c} \frac{\partial^2 u_i}{\partial x_i x_j} - \mathcal{H}(u_i),
    \label{eq:incompressibe_N-S}
\end{align}
\noindent where the usual index notation is used, 
$u_i$ denotes the velocity components in Cartesian coordinates, 
$p$ denotes the pressure, 
and $\mathcal{H}$ is the high-pass filter {ensuring that the LES is}  acting on a subset of modes within each spectral element. 
The implementation of this filter in Nek5000 has been validated by~\citet{vinuesa_turbulent_2018}.

The computational domain is a C-mesh extending $6c$ horizontally and $4c$ vertically. 
The leading and trailing edges of the airfoil are located $2c$ and $3c$ from the front boundary of the domain, respectively. 
The spanwise width of the domain for simulating the NACA4412 wing section is $0.2c$ and for the NACA0012 wing section is $0.1c$. 
This setup has been validated {in} previous works~\citep{vinuesa_turbulent_2018,tanarro_effect_2020,atzori_uniform_2021}, 
confirming that the spanwise widths are sufficient to capture the range of active scales 
as well as the largest turbulent structures in the TBL over the airfoil. 
The computational domain for the NACA4412 and NACA0012 consists of 
$127 \ 000$ and $220 \ 000$ spectral elements, respectively.

The desired near-wall spatial resolution of the current simulation is expressed 
in terms of viscous units such that {$\Delta x^+_t < 18.0$, $\Delta y^+_n < (0.64, 11.0)$, and $\Delta z^+ < 9.0$} 
in the wall-tangential, wall-normal, and spanwise directions, respectively. 
{To achieve this resolution, {polynomial orders} of $P = 11$ and $P = 7$ are adopted {for simulating the flow around} the NACA4412 and NACA0012 {cases}, respectively. 
This leads to total number of grid points of $\approx 1.1 \times 10^{8}$ and $\approx 2.2 \times 10^{8}$ for NACA0012 and NACA412 cases, respectively.} 
{Note that the} viscous length $l^* = \nu / u_{\tau}$ is defined in terms of the {friction} velocity $u_{\tau} = \sqrt{\tau_{w} / \rho}$ 
(where $\tau_{w} = \rho \nu {\left ( {\rm d} U_t / {\rm d} y_n \right )}_{ y_n = 0 }$ 
is the mean {wall-shear} stress and $\rho$ is the density) and fluid kinematic viscosity $\nu$. 
The spatial resolution in the wake follows $\Delta_x / \eta < 9$, 
where $\eta = {(\nu^3/\epsilon)}^{1/4}$ represents the Kolmogorov scale. 
Note that $\epsilon$ is the local isotropic dissipation. 
The proposed mesh has been validated for {the present SGS approach} by~\citet{negi_unsteady_2018}, 
demonstrating a very good agreement between DNS and LES results for a simulation of a NACA4412 airfoil at ${Re}_c = 400 \ 000$.

Dirichlet boundary conditions {were} adopted for the front, upper, and lower boundaries, 
imposing an estimated far-field velocity distribution obtained via a {supporting} Reynolds-averaged {Navier--Stokes} (RANS) simulation. 
The $k$--$\omega$ {shear-stress transport (SST)} model~\citep{menter_kmSST_1994} {was} employed for the RANS simulation, 
where a circular domain with {a radius of} $200c$ is considered. 
For the outlet, the boundary condition developed by~\citet{dong_robustoutlet_2014} is used, 
preventing the uncontrolled influx of kinetic energy through the outflow boundaries. 
The boundary layer is tripped at $x/c = 0.1$ from the leading edge on both suction and pressure sides of the wing section. 
The tripping {consists of} a wall-normal localized body force, {imitating the effect of experimental devices~\citep{hosseini_DNSwing_2016}.
}

{The simulation procedure is as follows}: 
(1) Use the solution of the RANS simulation as the initial condition for velocity and pressure fields. 
(2) Run the simulation at a polynomial order of $P = 5$ for $6$ flow-over times, and then for $10$ flow-over times at a polynomial order of $P = 7$. 
(3) Run the simulation at the target polynomial order of $P=11$ for $11$ time units 
{while accumulating samples to compute} the statistics. 
{This process ensures a smooth increase of resolution from the initial condition, 
as well as the identification and removal of initial transients~\citep{vinuesa_wingTBL_2017}.} 
Note that one flow-over time is defined as the time required 
for a fluid particle moving with the incoming velocity $U_{\infty}$ to travel a distance of $c$. 
For simulating uncontrolled NACA0012 cases, we also use the solution of the RANS simulation as the initial condition, 
{and} directly start running the simulation at a polynomial order of $P = 7$ for $7$ flow-over times, 
and subsequently run for $20$ flow-over times for sampling the statistics. 
 Running {for} $20$ flow-over times is {necessary} due to 
fact that the {spanwise width} of the NACA0012 wing section is half of that of the NACA4412 in the present study{, see~\citet{vinuesa_ductconvergence_2016,vinuesa_turbulent_2018} for additional details}. 
For the cases with the control scheme applied, 
we start the simulation from the fully-developed turbulent field of the uncontrolled case. 
Each controlled case on the airfoil NACA4412 and NACA0012 requires approximately $10$ and $20$ flow-over times, respectively, to obtain converged statistics. 
The computational cost to simulate $10$ flow-over times is approximately 1 million CPU hours on a Cray-XC 40 system.

\subsection{Control schemes and configurations}\label{sec:control_scheme}

In the present study, we investigate five different control cases, whose configurations are summarized in table~\ref{tab:ctrl_configs}. 
In cases OC1 and BL1, opposition control and uniform blowing are applied on the suction side of the NACA0012 airfoil 
between the streamwise locations of $x/c = 0.25$ and $0.86$, following the setup by~\citet{vinuesa_skin_2017}. 
Cases OC2 and BL2 involve OC and uniform blowing applied to the suction side of the NACA4412 airfoil within the same streamwise range. 
Additionally, we include a case BD2 from the work of~\citet{atzori_uniform_2021}, 
which applies body-force damping to the suction side of the NACA4412 airfoil within the same streamwise range.

\begin{table}
    \begin{center}
        \def~{\hphantom{0}}
        \begin{tabular}{ccccc}
            Case notation  & Airfoil  &   $AoA$   & Control over suction side ($0.25 < x/c < 0.86$)    &   Color code           \\[3pt]
            ${\rm Ref1}$   & NACA0012 &   $0^{\circ}$   & Uncontrolled                                       &  \colorbox{myblack}{\makebox[0.1em][c]{\rule{0pt}{0.1em}}} \\
            ${\rm BL1} $   & $-$        &   $-$   & Uniform blowing    with $0.1\%U_{\infty}$          &  \colorbox{myblue}{\makebox[0.1em][c]{\rule{0pt}{0.1em}}}             \\
            ${\rm OC1} $   & $-$        &   $-$   & Opposition control with $y^+_s = 15$               &  \colorbox{myred}{\makebox[0.1em][c]{\rule{0pt}{0.1em}}}               \\ [5pt]

            ${\rm Ref2}$   & NACA4412 &  $5^{\circ}$  & Uncontrolled                                       &  \colorbox{myblack}{\makebox[0.1em][c]{\rule{0pt}{0.1em}}}                     \\
            ${\rm BL2} $   & $-$        &   $-$   & Uniform blowing    with $0.1\%U_{\infty}$          &  \colorbox{myblue}{\makebox[0.1em][c]{\rule{0pt}{0.1em}}}             \\
            ${\rm OC2} $   & $-$        &   $-$   & Opposition control with $y^+_s = 15$               &  \colorbox{myred}{\makebox[0.1em][c]{\rule{0pt}{0.1em}}}               \\ 
            ${\rm BD2} $   & $-$        &   $-$   & Body-force damping with $y^+_n < 20$               &  \colorbox{myyellow}{\makebox[0.1em][c]{\rule{0pt}{0.1em}}}               \\ 
        \end{tabular}
        \caption{Control configurations considered in the present study at ${Re}_c = 200 \ 000$. 
                The colored squares denote the color code for each case.}
        \label{tab:ctrl_configs}
        \end{center}
\end{table}

\begin{figure}
    \centerline{\includegraphics[width=\textwidth]{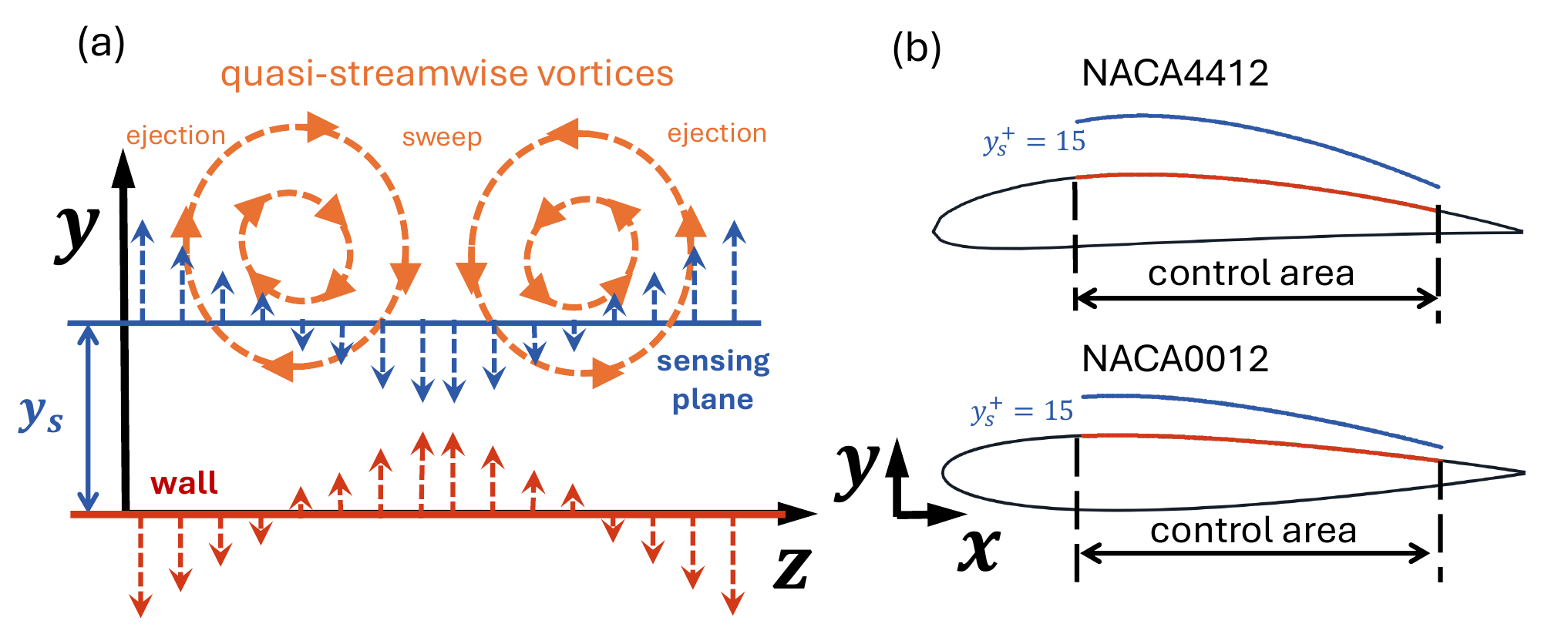}}
    \caption{(a) Schematic view {of the} opposition control scheme{, adapted from~\citet{stroh_comparison_2015}.} 
    (b){Schematic} of control configurations for NACA4412 (upper) and NACA0012 (lower).
    }
    \label{fig:oppoctrl_schematic}
\end{figure}

Opposition control is implemented as a Dirichlet boundary condition 
that imposes a velocity at the wall opposite to the detected wall-normal velocity at a prescribed sensing plane $y_s$, 
thereby {attenuating} sweep and ejection events near the wall~\citep{choi_active_1994}. 
{A} schematic {of the} OC mechanism is {shown} in figure~\ref{fig:oppoctrl_schematic}(a). 
This control method, also known as \textit{v-}control as defined by~\citet{choi_active_1994}, involves the control input at the wall given by
\begin{equation}
v(x,0,z,t) = -\alpha \left[ v(x,y_s,z,t) - \left\langle v(x,y_s,t) \right\rangle \right],
\label{eq:oppo_input}
\end{equation}
where $\alpha$ is a positive amplification factor set to 1 in the present study. 
Note that the spatial mean $\langle v(x,y_s,t) \rangle$ is subtracted to 
maintain a zero-net-mass-flux condition at the wall. 
Due to inherent streamwise nonhomogeneity, 
the spatial mean is obtained by averaging the instantaneous wall-normal velocity component in the spanwise direction.

The sensing plane $y_s$ is set to $y^{+}_{s} = 15$ based on previous studies~\citep{hammond_observed_1998}, 
which showed that the maximum drag reduction rate was achieved at this position in viscous units. 
The {wall-normal location} in {outer} units is calculated as $y_s = \nu y^{+}_{s} / u_{\tau}$, where $u_{\tau}$ is the friction velocity of the uncontrolled case. 
As the local friction velocity $u_{\tau}$ varies{ with $x$}, 
$y_s$ is nonuniform to maintain the same dimensionless distance from the wall, 
as shown in figure~\ref{fig:oppoctrl_schematic}(b).

We performed a series of tests on the sensing plane location from $y^+_s = 7.5$ to $17.5$, 
showing that positions below $12.5$ and above $15.0$ increase skin-friction drag, 
while finer calibrations between $12.5 \leq y^+_s \leq 15.0$ yield little improvement. 
We also explored applying OC on both the suction and pressure sides of NACA4412 
but found {negligible} improvement for drag reduction. 
Thus, we focus on applying control over the suction side in the present study.

To validate our code implementation, we applied OC to TCF at $Re_{\tau} = 180$ and $y^+_{s} = 15$ using DNS in the same solver. 
In TCF, OC is applied on both walls, with the spatial average in equation~(\ref{eq:oppo_input}) computed 
in the streamwise and spanwise directions due to homogeneity and periodicity~\citep{stroh_comparison_2015}.
{This leads to zero for the spatial average as expected in TCF.} 
The computational domain {had the following dimensions:} $\Omega = L_x \times L_y \times L_z = 10h \times 2h \times 5h$ 
(where $h$ is the channel {half height}){, with} $N_x \times N_y \times N_z = 270 \times 144 \times 270$ 
{grid points} and a polynomial order of $11$, 
at a Reynolds number based on {bulk velocity $U_{b}$ of ${Re}_{b} = 2 \ 800$. }
Turbulence statistics were computed for about $600$ eddy turnover times after the controlled flow was fully developed. 
Our results, showing a drag reduction rate of $22.46\%$, 
agree closely with the $22.96\%$ reported by~\citet{choi_active_1994}{, who} {used a} similar numerical setup with same control configuration, 
{thus validating} our OC implementation.

Uniform blowing is implemented as a boundary condition of Dirichlet type at the wall, 
imposing the horizontal and vertical velocity components such that the magnitude of wall-normal velocity 
equals {the desired} control intensity~\citep{atzori_uniform_2021}. 
A constant intensity of $0.1\% U_{\infty}$ {was} adopted~\citep{atzori_aerodynamic_2020}.

Body-force damping imposes a volume force defined as $g(v_n) = -\gamma (v_{n,x}, v_{n,y}, 0)$ 
within the wall-normal region $y^{+}_{n} < 20$, 
where $v_{n,x}$ and $v_{n,y}$ are the projections of 
the wall-normal component of the instantaneous velocity on the horizontal and vertical directions, respectively, 
and $\gamma$ is the coefficient of the {body-force} intensity~\citep{stroh_global_2016}. 
The momentum equation incorporating this volume force is expressed as: 
\begin{align}
    \frac{\partial u_i}{\partial t} + u_j \frac{\partial u_i}{\partial x_j} &= 
    -\frac{\partial p}{\partial x_i} + \frac{1}{{Re}_c} \frac{\partial^2 u_i}{\partial x_i x_j} - \mathcal{H}(u_i)
    + g(v_n).
    \label{eq:body_force}
\end{align}
Note that the $\gamma$ value is calibrated to $\gamma = 32.5$ to {achieve a} skin-friction reduction comparable {to that of} 
uniform blowing with an intensity of $0.1\% U_{\infty}$ 
between $x/c \approx 0.45$ {and} $0.8$~\citep{atzori_uniform_2021}. 
{This control scheme is included to allow a comparison between different closed-loop control strategies.}


\section{Control effects on integral quantities}\label{sec:bl_develop}
In this section, we first assess the streamwise development {of the TBL} in terms of integral quantities.  
Subsequently, based on the assessments, we evaluate the control efficiency for drag reduction. 

\subsection{Streamwise development of the boundary layer}
\begin{figure}
    \centering{
        \includegraphics[width=0.7\textwidth]{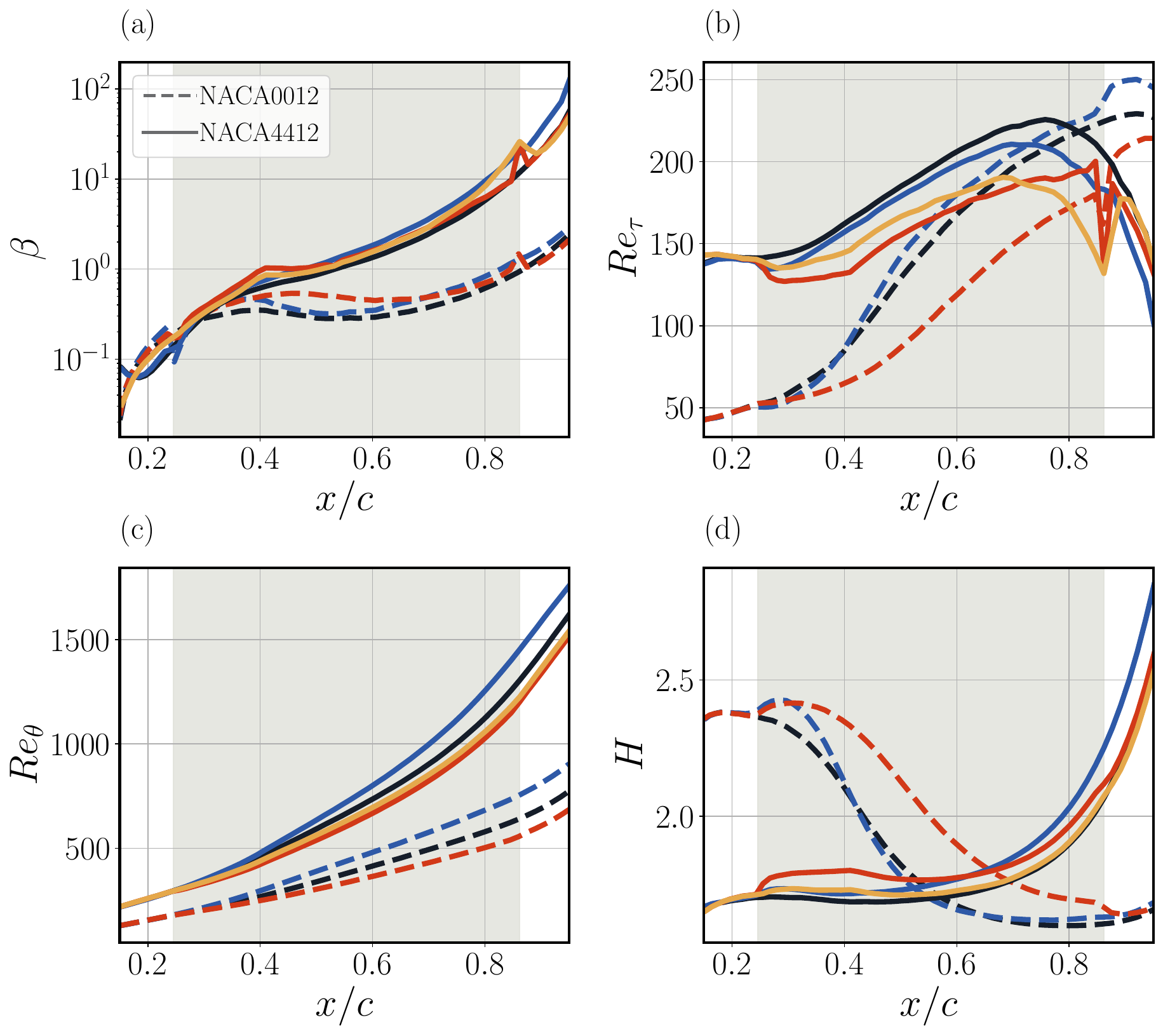}
    }
    \caption{
            (a) Clauser {pressure-gradient} parameter ($\beta$),
            (b) friction Reynolds number (${Re}_{\tau}$),
            (c) Momentum-thickness-based Reynolds number (${Re}_{\theta}$),
            and  
            (d) shape factor $H$ distributions on suction side of NACA0012 (dashed lines) and NACA4412 (solid lines).
            Note that the control region is indicated {in} gray.  
            The color code follows table~\ref{tab:ctrl_configs}.
            }
    \label{fig:BL_develop}
\end{figure}

The nonuniform {APGs} {in the various cases} {lead to} different interactions {with the} control schemes. 
We consider four parameters, namely the Clauser {pressure-gradient} parameter {$\beta = \delta^*/\tau_{w} {\rm d} P_e/ {\rm d} x_t$} 
(where $\delta^*$ is the displacement thickness and $P_e$ is the pressure at the boundary-layer edge), 
the friction Reynolds number ${Re}_{\tau} = \delta_{99} u_{\tau} / \nu$, 
the {momentum-thickness-based} Reynolds number ${Re}_{\theta} = U_e \theta / \nu$ and 
the shape factor $H = \delta^* / \theta$.
The streamwise {evolutions} are depicted in figure~\ref{fig:BL_develop}, 
and table~\ref{tab:BL_quantities} reports the parameters obtained at $x/c = 0.4$ and $0.75$.
Note that $\delta_{99}$ and $U_e$ are the $99\%$ boundary-layer thickness and the mean velocity at the boundary-layer edge, 
which are both obtained by the {method} proposed by~\citet{vinuesa_determining_2016}.

\begin{table}
    \begin{center}
        \def~{\hphantom{0}}
          \begin{tabular}{ccccccccccc}
            Case        & $x/c$  & $u_{\tau}$& $\theta$ & $\delta^*$ & $\delta_{99}$ & $\beta$ & $Re_{\tau}$ & $Re_{\theta}$ & $H$&  \\ 
            {\rm Ref1}  & $0.40$ & $0.0459$  & $0.0012$ & $0.0025$   & $0.0097$   & $0.35$ & $88$    &      $275$      & $2.1$ &  \\  
            {\rm BL1}   & $-$    & $0.0439$  & $0.0014$ & $0.0028$   & $0.0104$   & $0.45$ & $91$    &      $305$      & $2.1$ &  \\ 
            {\rm OC1}   & $-$    & $0.0376$  & $0.0011$ & $0.0026$   & $0.0088$   & $0.51$ & $66$    &      $254$      & $2.3$ &  \\ [2pt]
            {\rm Ref2}  & $0.40$ & $0.0635$  & $0.0018$ & $0.0029$   & $0.0130$   & $0.64$ & $164$   &      $470$      & $1.7$ &  \\ 
            {\rm BL2}   & $-$    & $0.0599$  & $0.0018$ & $0.0031$   & $0.0133$   & $0.76$ & $159$   &      $494$      & $1.7$ &  \\ 
            {\rm OC2}   & $-$    & $0.0522$  & $0.0016$ & $0.0029$   & $0.0127$   & $1.03$ & $132$   &      $435$      & $1.8$ &  \\  
            {\rm BD2}   & $-$    & $0.0571$  & $0.0017$ & $0.0029$   & $0.0128$   & $0.86$ & $146$   &      $452$      & $1.7$ &  \\ [6pt] 
            {\rm Ref1}  & $0.75$ & $0.0507$  & $0.0025$ & $0.0041$   & $0.0202$   & $0.45$ & $205$   &      $527$      & $1.6$ &  \\  
            {\rm BL1}   & $-$    & $0.0470$  & $0.0030$ & $0.0048$   & $0.0226$   & $0.60$ & $213$   &      $617$      & $1.6$ &  \\ 
            {\rm OC1}   & $-$    & $0.0447$  & $0.0022$ & $0.0038$   & $0.0179$   & $0.58$ & $160$   &      $459$      & $1.7$ &  \\ [2pt]
            {\rm Ref2}  & $0.75$ & $0.0417$  & $0.0043$ & $0.0078$   & $0.0270$   & $3.40$ & $225$   &      $988$      & $1.8$ &  \\ 
            {\rm BL2}   & $-$    & $0.0358$  & $0.0048$ & $0.0092$   & $0.0294$   & $5.18$ & $210$   &      $1096$     & $1.9$ &  \\ 
            {\rm OC2}   & $-$    & $0.0376$  & $0.0040$ & $0.0074$   & $0.0252$   & $3.93$ & $189$   &      $902$      & $1.8$ &  \\ 
            {\rm BD2}   & $-$    & $0.0356$  & $0.0041$ & $0.0074$   & $0.0259$   & $4.43$ & $184$   &      $931$      & $1.8$ &  \\ [6pt] 
            
          \end{tabular}
          \caption{{Integral} quantities at $x/c = 0.4$ and $0.75$.}
          \label{tab:BL_quantities}
          \end{center}
\end{table}

Figure~\ref{fig:BL_develop}(a) shows the streamwise evolution of the Clauser pressure-gradient parameter along {the suction} sides {of the} wing sections. 
It can be observed that the $\beta$ {curve of} case {\rm Ref2} is more prominent with respect to {\rm Ref1} and exhibits a rapid increase, 
which is attributed to the presence of the camber and higher angle of attack~\citep{tanarro_effect_2020}. 
Control schemes generally increase $\beta$ within the control region. 
OC and body-force damping significantly reduce $\tau_w$, enhancing $\beta$, 
whereas uniform blowing, though less effective in reducing $\tau_w$, increases $\delta^*$ more notably.
It can be observed that uniform blowing extends the increment {beyond the control region},  
while the reactive schemes recover $\beta$ to the uncontrolled states outside the control region.
Note that the profiles of OC and body-force damping exhibit a peak at the edge {of the} control area, 
reflecting strong {wall-pressure} fluctuations triggered {by the} control activations~\citep{stroh_comparison_2015}.

Figure~\ref{fig:BL_develop}(b) illustrates the distribution of ${Re}_{\tau}$ along the suction side, 
with case Ref1 showing continuous growth, and case Ref2 saturating before declining under the influence of increasingly intense pressure gradient. 
OC significantly reduces ${Re}_{\tau}$ by effectively lowering $u_{\tau}$, 
which is mirrored by body-force damping. 
In contrast, uniform blowing, by {increasing} $\delta^*$, 
leads to less reductions in ${Re}_{\tau}$, 
and may even increase it as observed in case BL1.

The evolution of ${Re}_{\theta}$, as shown in figure~\ref{fig:BL_develop}(c), 
is notably affected {by the} APG, which decelerates the flow and {increases the} momentum thickness $\theta$. 
Both OC and body-force damping counteract these effects, whereas uniform blowing exacerbates them. 
The contrasting impacts of these control strategies are significant, 
as the momentum thickness is closely related to skin friction at the wall. 
Note that the concept of virtual origin outlined by~\citet{stroh_global_2016} to describe the downstream control effect 
is difficult to evaluate for APG TBL due to the complex implication of flow history.

Finally, the evolution of the shape factor $H$ is depicted in figure~\ref{fig:BL_develop}(d). 
It can be observed that case Ref2 shows a remarkable increase throughout, 
while case Ref1 reaches a plateau and then decreases, suggesting a shift towards a more turbulent state. 
A higher $H$ is indicative of a laminar-like state {(or a flow with decreased near-wall fluctuations as in large APGs)}, 
with values around 2.6 typical for Blasius boundary layers, 
whereas turbulent states in ZPG TBL usually range from 1.2 to 1.6~\citep{schlatter_assessment_2010}.


\subsection{Control efficiency}
\label{sec:drag_reduction}
Based on the assessment of streamwise development,   
in this section, we first describe the how control schemes affect skin friction, 
and then demonstrate the achieved aerodynamic efficiencies {as they are applied}.

{To evaluate the modification of friction drag we examine} the skin-friction coefficient $c_f$,  
which is defined as $c_f = \tau_{w} / (\frac{1}{2} \rho U^2_{e} )$.  
{The} drag-reduction rate $R$, 
according to the metrics proposed by~\citet{kametani_effect_2015}, 
is defined as
\begin{equation}
    R = 1 - \frac{c_f(x)}{c^{\rm ref}_{f}(x)},
    \label{eq:drag_reduction}
\end{equation}
\noindent where superscript ${\rm ref}$ denotes the uncontrolled case. 
In addition to the drag reduction, the net-energy saving $E(x)$ is evaluated as 
\begin{equation}
    E(x) = 1 - \frac{ \left (c_f(x) + \psi_{\rm in}(x) \right )}{c^{\rm ref}_{f}(x)},
    \label{eq:energy_saving}
\end{equation}
\noindent where $\psi_{\rm in}(x)$ is the input power calculated as 
$\psi_{\rm in}(x) = \frac{1}{2} || v_{\rm wall} ||^3$ according to the approach of \citet{kametani_effect_2015}.
In the present study, the input power for OC is $\psi_{\rm in} \sim \mathcal{O}(10^{-6})$ 
while {in the case of} uniform blowing {it is} $\psi_{\rm in} \sim \mathcal{O}(10^{-12})$. 
The skin-friction coefficient is around $\mathcal{O}(10^{-3})$, 
suggesting that $R \approx E$ as discussed by~\citet{kametani_effect_2015}. 
However, this estimation is not {meaningful} for body-force damping 
{due to the fact that there is no actuator associated with this theortical control scheme.}

\begin{figure}
    \centerline{\includegraphics[width=\textwidth]{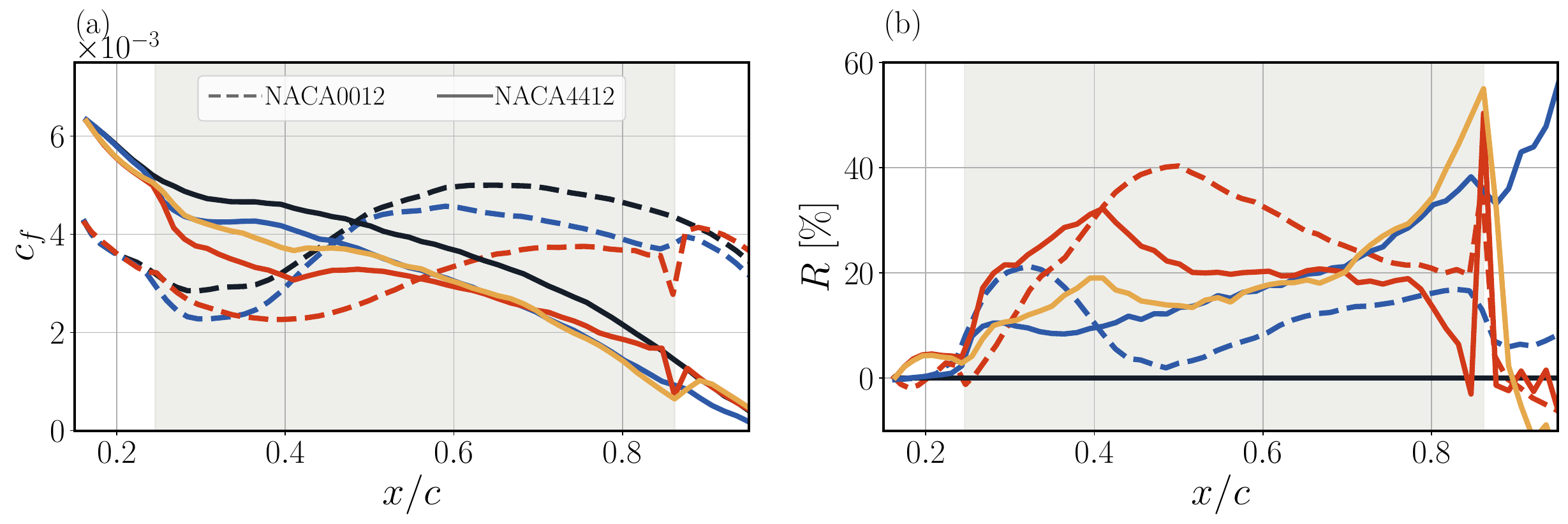}}
    \caption{(a) Skin-friction coefficient ($c_f$) over the suction side of NACA0012 (dashed lines) and NACA4412 (solid lines).
            (b) Spatial development of {drag-reduction} rate($R$) 
            over the suction side of NACA0012 (dashed lines) and NACA4412 (solid lines). 
            Note that the control region is colored {in} gray.  
            The color code follows table~\ref{tab:ctrl_configs}.
            }
    \label{fig:Cf_and_R_and_E}
\end{figure}

Figure~\ref{fig:Cf_and_R_and_E}(a) illustrates the streamwise development of $c_f$ 
along the suction sides of the wing sections. 
The variation in $c_f$ correlates closely with the evolution of the Clauser {pressure-gradient} parameter $\beta$.  
In particular, case Ref1 exhibits a peak in $c_f$ around $x/c \approx 0.6$, 
then gradually decreases due {to the} milder APG in NACA0012. 
Conversely, $c_f$ for case Ref2 sharply declines towards the trailing edge due {to the} stronger APG.
This leads to different {performances} of a control scheme applied on the wing sections, 
which are depicted by the obtained drag-reduction {rates} in figure~\ref{fig:Cf_and_R_and_E}(b).

OC results in a variable drag-reduction rate, reaching up to $40\%$ 
at $x/c \approx 0.5$ for case OC1 and approximately $30\%$ at $x/c = 0.4$ for case OC2. 
Note that the saturation in OC2 is connected with the presence of maximum camber at the same streamwise location. 
Beyond the saturation point and towards {the end of the control region}, case OC1 maintains about $R \approx 20\%$, 
whereas OC2 experiences a reduction at $x/c \approx 0.85$.
This is due to the inherently intensifying APG in this area, {resulting in values of $\beta$} of an order of magnitude higher {for case {\rm Ref2}} than case {\rm Ref1}.
Note that the peaks at the edge of {the control region} {produced} by {\rm OC1} and {\rm OC2} are $\approx 38\%$ and $\approx 50\%$, respectively{,}
aligning with the aforementioned strong pressure fluctuations caused {by the control.}

{The other} reactive scheme, BD2 does not achieve the same saturation as OC but instead shows a peak of about $20\%$ at $x/c = 0.4${,} 
and a constant increase in {the regions} with strong APG {is} observed. 
This difference is attributed to the calibrated amplitude $\gamma$. 
Additionally, case BD2 also peaks at the edge of control with similar value of the case {\rm OC2}.

{The performance of uniform blowing varies {depending on the wing case under study.}} 
In particular, case BL2 shows a continuous growth from {$R \approx 10\%$} up to $40\%$, 
corresponding to a nearly constant absolute reduction in $c_f$~\citep{atzori_aerodynamic_2020}.
In contrast, case ${\rm BL1}$ exhibits a saturated state of $R \approx 20\%$ at $x/c = 0.3$, which corresponds to the location of {the max thickness of the airfoil}. 
In regions of stronger APG, it produces a minimum of $\approx 2\%$ at $x/c \approx 0.5$, followed by continuous growth similar to ${\rm BL2}$.

It is worth noting that, {downstream of the control region}, 
uniform blowing results in positive $R$ values,  
whereas reactive {schemes} generally show negative outcomes,  
which are aligned with the findings from previous studies~\citep{stroh_comparison_2015,stroh_global_2016,atzori_aerodynamic_2020}.
Additionally, over the pressure side, none of the considered control schemes impact $c_f$. 
Note that {an additional} {assessment of the} skin-friction coefficient {using the} FIK identity is reported in appendix~\ref{sec:FIK}.

Next, we evaluate the control effects on the aerodynamic efficiency (\textit{i.e.} lift-to-drag ratio) of the airfoil, 
which is typically expressed as the ratio between lift and drag coefficients. 
The lift and drag coefficients are defined as {$C_l = f_l/(bq)$} and {$C_d = f_d/(bq)$}, respectively, 
where we denote {$b$ as spanwise width, whereas} {$f_l/b$} as the lift force per unit length, {$f_d/b$} as the drag force per unit length 
and $q$ as the free-stream dynamic pressure ($q = \frac{1}{2} \rho U^2_{\infty} $).
Note that the {$f_l/b$} and {$f_d/b$} are calculated by integrating wall-shear stress and the pressure force 
{around the wing section and projecting them on} the directions parallel and perpendicular to the incoming flow, respectively. 
To investigate the contributions {of the} drag coefficient{$C_d$}, 
we further decompose {it} into {skin-friction} drag 
{$C_{d,f} = f_{d,f}/(bq)$} and pressure drag {$C_{d,p} = f_{d,p}/(bq)$}.

Table~\ref{tab:ClCd} summarizes the values {of the} integrated lift ($C_l$), 
integrated skin-friction ($C_{d,f}$), 
and pressure ($C_{d,p}$) contributions to the total drag ($C_d$), 
and the aerodynamic efficiency {($L/D = C_l / C_d$)} for the cases considered in the present study.
{Note that as the $C_l$ for case {\rm Ref1} is approximately $0$ and thus $L/D \approx 0$, 
we focus on the modification of $C_d$ components for cases {\rm OC1} and {\rm BL1}.}
Uniform blowing applied on the suction side 
reduces the friction drag force thus reduces the $C_{d,f}$ 
while {increasing} the $C_{d,p}$, which is also reported by~\citet{atzori_aerodynamic_2020}. 
Consequently, case {\rm BL1} does not change the $C_d$ and case {\rm BL2} increase the $C_d$ by $2.8\%$ with respect to the uncontrolled case.
However, applying OC on the suction side (case {\rm OC1} and {\rm OC2}) reduces both $C_{d,f}$ and $C_{d,p}$ simultaneously. 
The resulting reductions in $C_d$ are $6.8\%$ and $5.1\%$ for case {\rm OC1} and {\rm OC2}, respectively, 
being similar to the effects of applying body-force damping on the suction side.
\begin{table}
    \begin{center}
        \def~{\hphantom{0}}
        \begin{tabular}{cccccc}
            Case           & $C_l$                                    &  $C_{d,f}$                            & $C_{d,p}$                               & $C_d = C_{d,f} + C_{d,p}$                     &  $L/D$     \\[3pt]
            ${\rm Ref1}$   & $\approx 0$                                &  0.0105                               &  0.0014                                 & 0.0118                                        &   $\approx 0$   \\
            ${\rm BL1} $   & $-$     &  0.0101 \textcolor{mygreen}{($-3.8\%$)} &  0.0017 \textcolor{red}{($+21.4\%$)}    & 0.0118 \textcolor{blue}{($+0.0\% $)}                                            &  $-$  \\
            ${\rm OC1} $   & $-$     &  0.0097 \textcolor{mygreen}{($-7.6\%$)} &  0.0014 \textcolor{blue}{($+0.0\%$)}    & 0.0110 \textcolor{mygreen}{($-6.8\% $)}                                           & $-$ \\ [5pt]
            ${\rm Ref2}$   & $ 0.8670$                                &  0.0128                               &  0.0087                                 & 0.0215                                        &   41   \\
            ${\rm BL2} $   & $ 0.8330$ \textcolor{red}{($-4.0\%$)}    &  0.0122 \textcolor{mygreen}{($-4.7\%$)} &  0.0099 \textcolor{red}{($+13.8\%$)}    & 0.0221 \textcolor{red}{($+2.8\% $)}           &   38 \textcolor{red}{($-7.3\%$)}  \\
            ${\rm OC2} $   & $ 0.8734$ \textcolor{mygreen}{($+1.0\%$)}    &  0.0121 \textcolor{mygreen}{($-5.5\%$)} &  0.0083 \textcolor{mygreen}{($-4.6\%$)}   & 0.0204 \textcolor{mygreen}{($-5.1\% $)}         &   43 \textcolor{mygreen}{($+4.9\%$)}  \\ 
            ${\rm BD2} $   & $ 0.8790$ \textcolor{mygreen}{($+1.4\%$)}    &  0.0121 \textcolor{mygreen}{($-5.5\%$)} &  0.0083 \textcolor{mygreen}{($-4.6\%$)}   & 0.0204 \textcolor{mygreen}{($-5.1\% $)}         &   43 \textcolor{mygreen}{($+4.9\%$)}  \\ 
        \end{tabular}
        \caption{Integrated lift ($C_l$), integrated skin-friction ($C_{d,f}$), and pressure ($C_{d,p}$) contributions to the total drag ($C_d$), 
                and the aerodynamic efficiency ($L/D$) for the cases considered in the present study.
                The values in the parentheses report the relative changes obtained by control.}
        \label{tab:ClCd}
        \end{center}
\end{table}

\section{Control effects on turbulence statistics}\label{sec:stat}
{The adverse} pressure gradient intensifies wall-normal convection, 
significantly affecting the outer region of the boundary layer~\citep{vinuesa_wingTBL_2017}. 
This has a substantial impact on the turbulence statistics, 
which is crucial for understanding the {different} efficiencies {of the various} {drag-reduction} schemes.

In this section, we {analyze} the mean flow, {Reynolds stress}, and turbulent kinetic energy (TKE) budgets. 
We assess {inner- and outer-scaled} profiles at $x/c = 0.4$ and $0.75$, 
corresponding to moderate and strong APG conditions, respectively. 
Comparing these profiles {helps to} distinguish control effects 
due to variations in friction velocity {and true profile} modification. 
Note that the statistics are expressed in terms of wall-tangential ($t$) and normal ($n$) directions 
relative to the wing surface, 
using the local quantities to determine the corresponding inner and outer scales for all cases. 

\subsection{Mean velocity profile}
\begin{figure}
    \centering{
        \includegraphics[width=\textwidth]{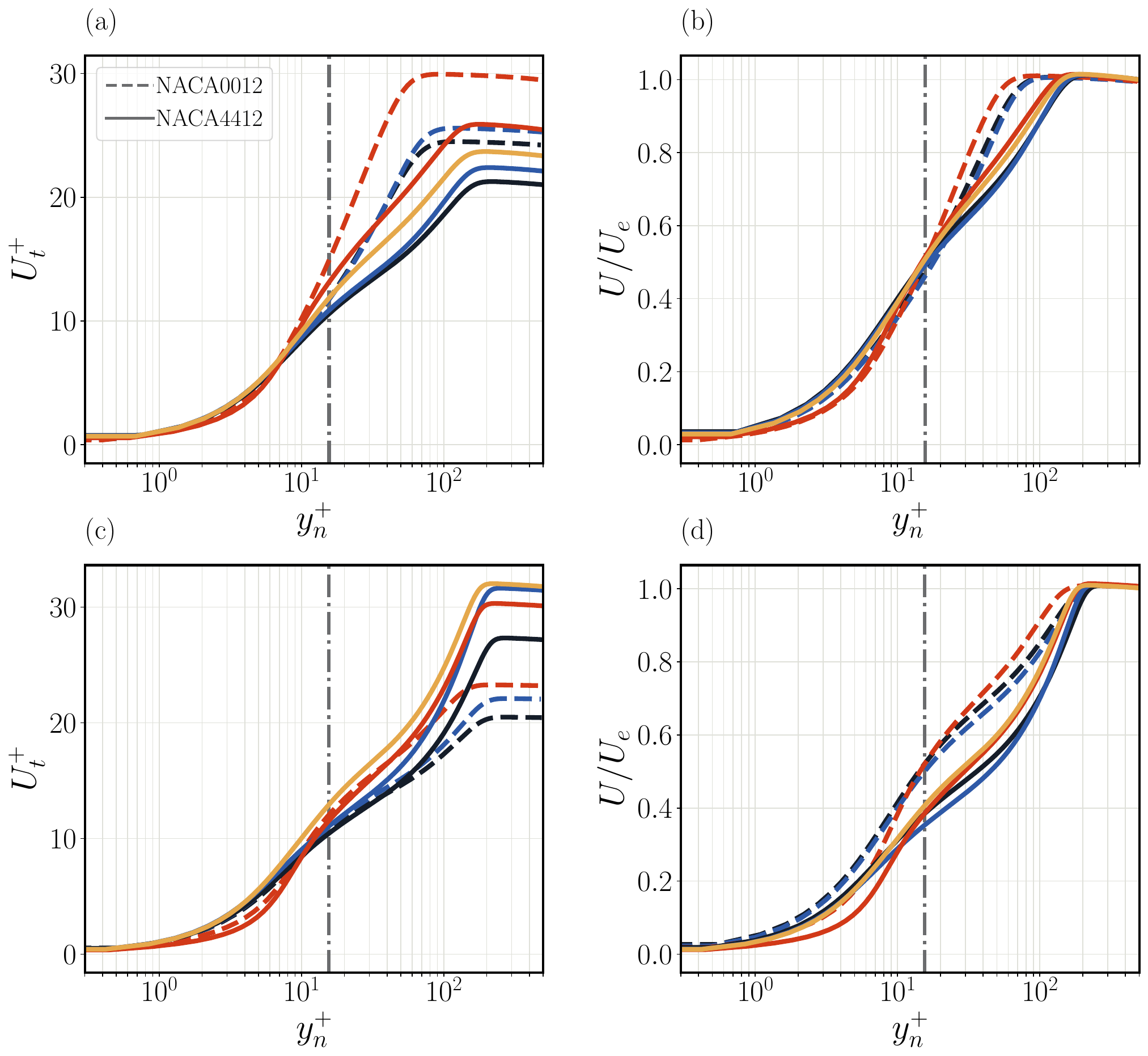}
    }
    \caption{
            (Left) Inner- and (right) outer-scaled mean wall-tangential velocity
            ${U_t}$ 
            as a function of the inner-scale wall-normal distance $y^+_{n}$ 
            at (top) $x/c = 0.4$ and (bottom) $x/c=0.75$  
            on suction side of NACA0012 (dashed lines) and NACA4412 (solid lines). 
            The prescribed sensing plane for OC of $y^+_{s} = 15$ is indicated by a gray dash-dotted line.          
            The color code follows table~\ref{tab:ctrl_configs}.
            }
    \label{fig:U_inner_040_075}
\end{figure}

Figure~\ref{fig:U_inner_040_075}(a) and~\ref{fig:U_inner_040_075}(c) depict the 
inner-scaled mean wall-tangential velocity profiles $U^{+}_t$ at $x/c = 0.4$ and $x/c = 0.75$ 
on the suction side of the wing sections, respectively.
Stronger {APGs} are known to enhance the wake region in inner-scaled profiles~\citep{spalart_experimental_1993,monty_APGparametric_2011}. 
\citet{tanarro_effect_2020} observed that {among the two uncontrolled cases}, 
the wake region was more pronounced at both locations {for the} NACA4412.

This effect remains evident {in the} uncontrolled cases at $x/c=0.75$, 
where case {\rm Ref2} exhibits a higher pressure gradient ($\beta = 3.4$) than {\rm Ref1} ($\beta = 0.45$). 
However, at $x/c=0.4$, {\rm Ref1} shows a more pronounced wake region 
despite a lower pressure gradient ($\beta = 0.35$) compared to {\rm Ref2} ($\beta = 0.64$). 
This is {in fact a low-Reynolds number effect as documented by~\citet{vinuesa_turbulent_2018}},  
and a higher shape factor {can be observed in} {\rm Ref1} at this location, 
as detailed in table~\ref{tab:BL_quantities}. 
{The Reynolds number is also too low to observe any significant overlap region.} 
Nonetheless, {at the downstream location} $x/c = 0.75$, 
the stronger APG results in a steeper slope in {the very incipient} overlap {region found there}~\citep{vinuesa_wingTBL_2017,tanarro_effect_2020}.


OC causes a significant downward shift in the viscous sublayer 
and an increase in $U^+_{t}$ {starting} from $y^{+}_{n} \approx 7$, 
becoming more pronounced with increasing APG intensity. 
\citet{choi_active_1994} {related} this shift {with} the displacement of the so-called virtual origin. 
However, due to the complex history effects of {the APGs studied here}, 
it remains challenging to conclusively evaluate this impact in the current study~\citep{atzori_uniform_2021}. 
The prominent wake region is further intensified by the opposition control.
In particular, the case {\rm OC1} results in the highest $U^+_t$ at $x/c = 0.4$ and $0.75$ 
while case {\rm OC2} {yields} slightly lower $U^+_t$ than {\rm BD2} at $x/c =0.75$ in the wake region and {farther} away.
This is due to the {different reduction} in $u_{\tau}$ at {the various} streamwise locations.

Uniform blowing {increases the inner-scaled} velocity in the wake region compared to the reference. 
In the buffer layer, this control scheme increases $U^+_t$ but to a less extent than OC, 
being similar to the effects observed under stronger APGs~\citep{atzori_uniform_2021}. 
The influence on the viscous sublayer is less {pronounced} than that of OC but is nonetheless significant.

Body-force damping impacts $U^+_t$ similarly to uniform blowing but exhibits additional complexity. 
For instance, case {\rm BD2} exhibits higher $U^+_t$ than {\rm BL2} across most wall-normal distances, 
despite achieving comparable friction-drag reduction at $x/c = 0.75$. 
This similarity extends to the {boundary-layer} edge, where it aligns closely with {\rm BL2}.

Control impacts on inner-scaled profiles are largely tied to modifications in friction velocity. 
We further illustrate this by presenting outer-scaled profiles $U_{t}/U_{e}$ in figure~\ref{fig:U_inner_040_075}(b) and (d) 
for $x/c = 0.4$ and $0.75$, respectively. 
Despite the {low-Reynolds-number} effect at $x/c=0.4$, 
the outer-scaled profile of case {\rm Ref1} significantly exceeds that of {\rm Ref2} 
in both the viscous sublayer and buffer layer at $x/c = 0.75$, 
where the flow experiences substantial deceleration.

OC profiles are notably shifted downward in the viscous sublayer, 
with this effect extending into the buffer layer. 
The controlled profiles subsequently realign with the reference case 
around the position of the prescribed sensing plane ($y^+_n = 15$). 
Beyond this wall-normal position, OC {increases} the outer-scaled mean velocity.

Body-force damping slightly reduces velocity in the viscous sublayer due to a decrease in friction velocity. 
Its effects above this layer {are} similar to those of OC, 
with case {\rm BD2} aligning closely with {\rm OC2} but showing slight variations in velocity at higher wall-normal distances.
 
Uniform blowing, in contrast, consistently leads to lower outer-scaled profiles than the uncontrolled cases, 
particularly under stronger APGs. 
This is primarily linked to reduced friction velocity.

\begin{figure}
    \centering{
        \includegraphics[width=0.85\textwidth]{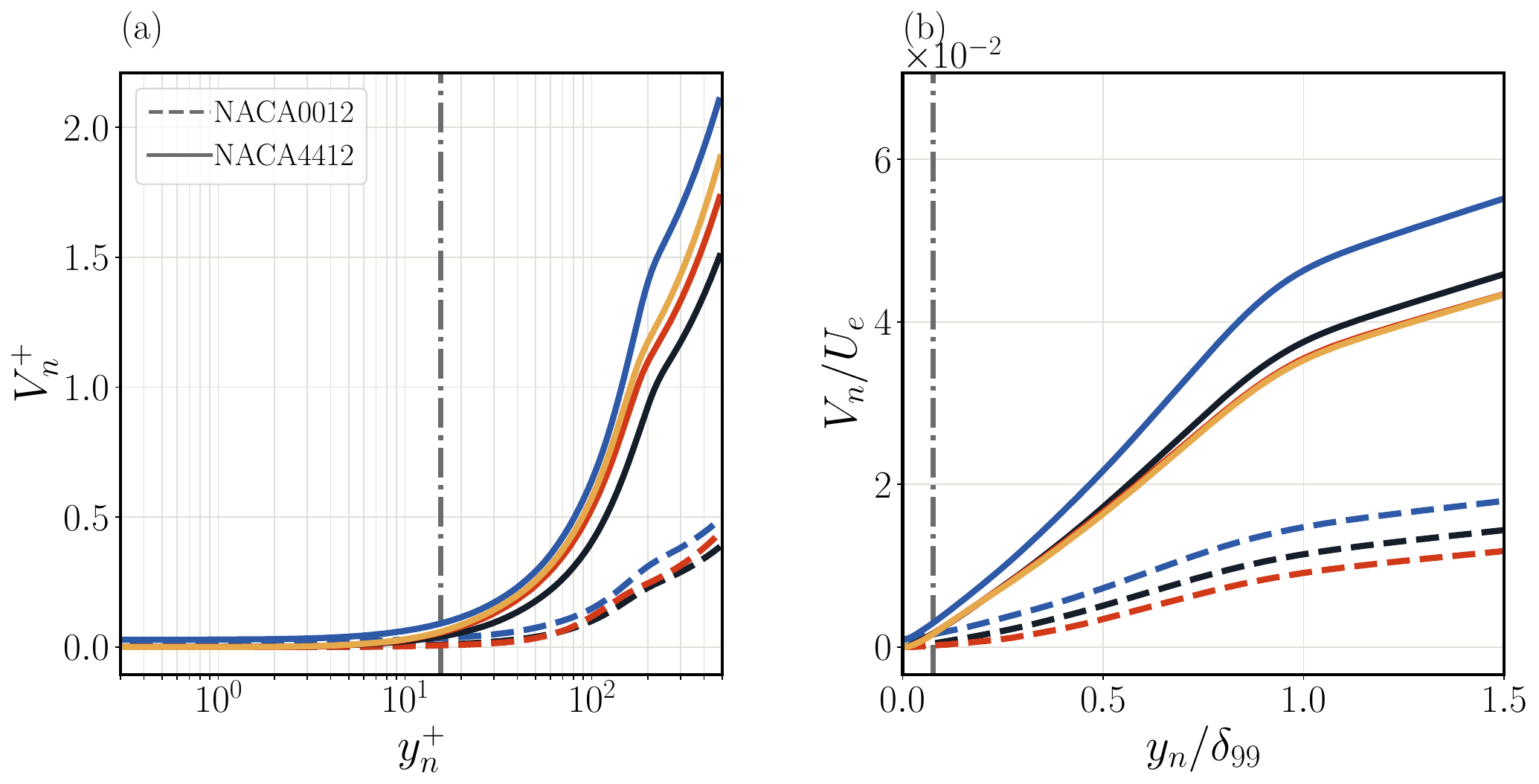}
    }
    \caption{
            (a) Inner- and (b) outer-scaled  mean wall-normal velocity components $V_n$
            as a function of the wall-normal distance $y_n$ at $x/c = 0.75$ 
            on suction side of NACA0012 (dashed lines) and NACA4412 (solid lines).
            The prescribed sensing plane of OC ($y^+_{s} = 15$) 
            corresponded to inner-scaled and outer-scaled positions
            are indicated by a gray dash-dotted line in (a) and (b), respectively.          
            The color code follows table~\ref{tab:ctrl_configs}.
            }
    \label{fig:V_inner_outer}
\end{figure}

Adverse pressure gradients (APGs) significantly {increase the} wall-normal convection, 
which enhances {energy in the outer region} of the boundary layer, 
impacting {the mean} wall-tangential velocity profiles~\citep{harun_PGEffect_2013,vinuesa_wingTBL_2017,vinuesa_turbulent_2018}. 
Consequently, we examine the inner-scaled mean wall-normal velocity component $V_n$ 
at $x/c = 0.75$ as shown in Figure~\ref{fig:V_inner_outer}(a). 
One can observe that case {\rm Ref2} exhibits a more pronounced $V_n$ across all wall-normal distances, 
{confirming} that stronger APGs further {increase} the mean wall-normal convection.

OC does not significantly alter the mean wall-normal convection. 
In particular, OC maintains the same $V^+_n$ values as the reference up to $y^+_n \geq 10$. 
Beyond this position, the profiles of {\rm OC1} and {\rm OC2} show divergent trends; 
{\rm OC1} shifts downward, while {\rm OC2} shows an upward shift compared to the uncontrolled cases. 
However, farther from the wall, both cases exhibit an increase in $V^+_n$, 
following the trend of the uncontrolled cases, 
with {\rm OC2} displaying a more pronounced increment.

The impact of body-force damping is very similar to that of opposition control. 
In particular, the profile of {\rm BD2} follows a similar pattern to {\rm OC2} 
but exhibits a greater increase in $V^+_n$ at wall-normal distances greater than $y^+_n > 10^2$.

{As expected}, uniform blowing significantly intensifies $V^+_n$. 
Specifically, the profiles of {\rm BL1} and {\rm BL2} show the highest $V^+_n$ values at all wall-normal distances 
on {both} wing sections. 
{For} $y^+_n < 10$, the $V^+_n$ matches the prescribed control intensity.

To isolate the effects of {friction-velocity} variations, 
we analyze the outer-scaled profiles $V_n / U_e$ at the same location, 
illustrated in figure~\ref{fig:V_inner_outer}(b). 
The comparison of profiles from uncontrolled cases indicates that 
stronger APGs significantly intensify mean wall-normal convection, 
being independent of {friction-velocity} modifications. 
The outer-scaled profiles of OC and body-force damping are slightly lower than those of the reference cases, 
despite the reduction in friction velocity, 
suggesting that these control methods do not directly 
{impact} mean wall-normal convection~\citep{atzori_uniform_2021}. 
In contrast, the outer-scaled profiles of uniform blowing are consistently higher than the uncontrolled cases.

\subsection{Reynolds-stress components}
The intensified wall-normal convection caused by APG significantly affects momentum transfer 
across the entire boundary layer~\citep{vinuesa_skin_2017}, 
which therefore impacts the wall-normal profiles of the {Reynolds-stress-tensor} components.

\begin{figure}
    \centering{
        \includegraphics[width=0.9\textwidth]{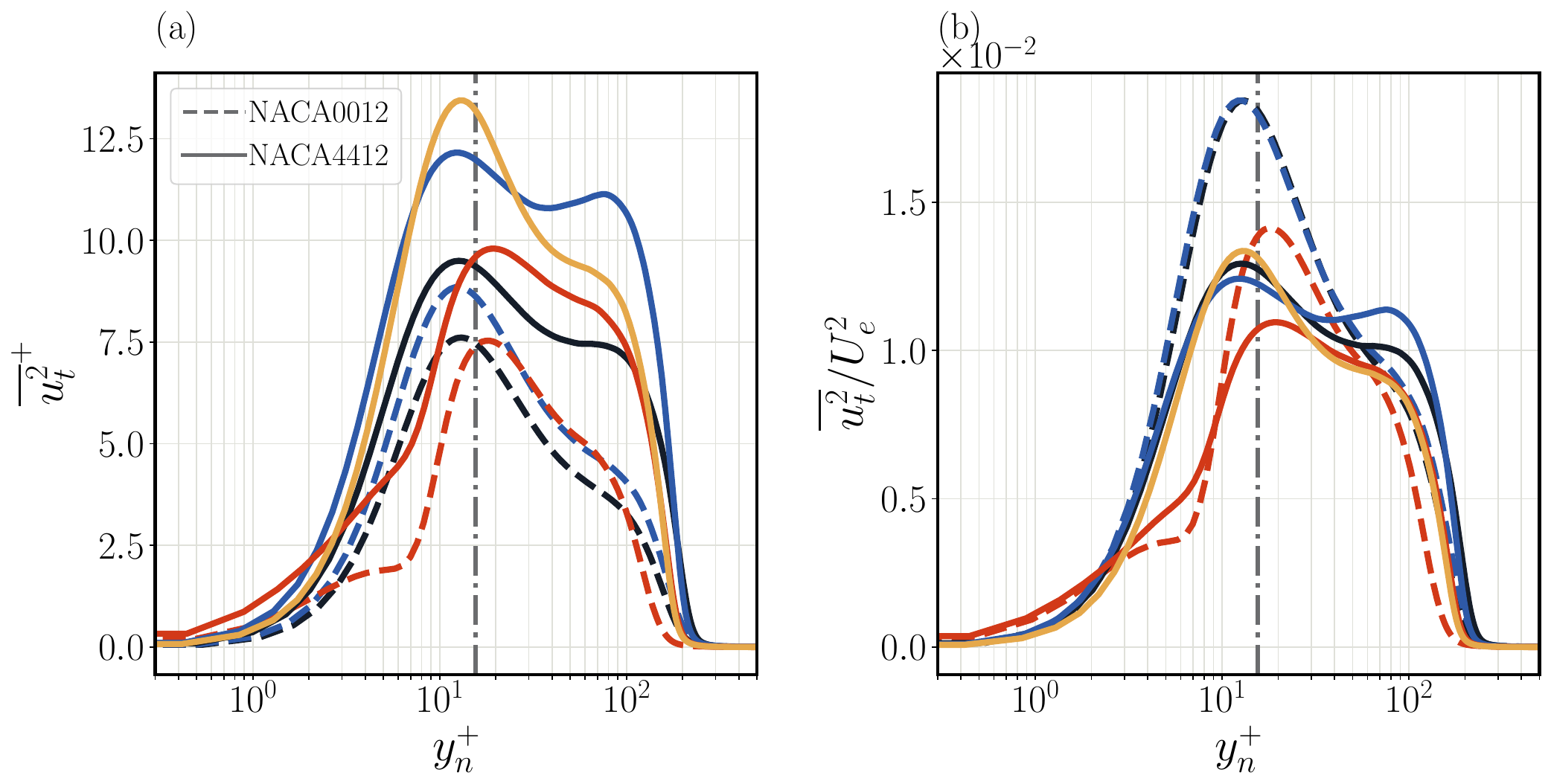}
    }
    \caption{(Left) Inner- and (right) outer-scaled fluctuations of wall-tangential velocity components 
            $\overline{u^2_t}$ 
            as a function of the inner-scale wall-normal distance $y^+_{n}$ at $x/c=0.75$  
            on suction side of NACA0012 (dashed lines) and NACA4412 (solid lines). 
            The dashed lines and solid lines {denote} the configurations for NACA0012 and NACA4412, respectively. 
            The prescribed sensing plane of OC ($y^+_{s} = 15$) 
            is indicated by a gray dash-dotted line.          
            The color code follows table~\ref{tab:ctrl_configs}.
            }
    \label{fig:uu}
\end{figure}

In figure~\ref{fig:uu}, we {show} the inner- and outer-scaled fluctuations of wall-tangential velocity components 
as a function of the inner-scaled wall-normal distance 
at {$x/c = 0.75$}. 
Regarding the inner-scaled profiles of case {\rm Ref2}, 
there is a more pronounced inner peak compared to case {\rm Ref1} at $x/c = 0.75$, 
indicating {that the} increased APG amplifies wall-tangential fluctuations at all wall-normal distances. 
This is connected to the lower friction velocity under intense APG conditions, 
as revealed by the outer-scaled profiles. 
Additionally, a noticeable effect {of the} stronger APG is the emergence of an outer peak~\citep{monty_APGparametric_2011} 
due to enhanced wall-normal convection.

For the OC cases, the inner-scaled profiles exhibit higher inner and outer peaks than the uncontrolled case, 
due to a reduction in friction velocity. 
The inner peak shifts {farther} from the wall while the outer peak descends towards the wall, 
both by approximately one wall unit. 
Furthermore, OC significantly attenuates wall-tangential fluctuations in the viscous sublayer, 
especially under milder adverse-pressure gradient conditions. 
Note that the increase in $\overline{u^2_t}$ near the wall is due to the {control input}~\citep{choi_active_1994}. 
Comparing the outer-scaled profiles, OC {yields} the greatest reduction in both the inner and outer peaks 
among the control schemes considered.

In contrast, uniform blowing further intensifies 
both the near-wall and outer peaks of $\overline{u^2_t}^+$ 
compared to the uncontrolled cases, 
which is similar to the effect of stronger APG. 
This is linked to a more significant reduction in friction velocity 
when the pressure gradient is less intense.

The effects of body-force damping on $\overline{u^2_t}^+$ and $\overline{u^2_t}$ 
differ qualitatively from those of uniform blowing and opposition control. 
In particular, the inner-scaled profile of case {\rm BD2} exhibits 
a higher inner peak and a lower outer peak compared to uniform blowing,
which is more prominent under stronger adverse pressure gradient. 
However, the inner peak of $\overline{u^2_t}$ is not attenuated, 
whereas the outer peak is lower than that of case {\rm BL2}.

\begin{figure}
    \centering{
        \includegraphics[width=\textwidth]{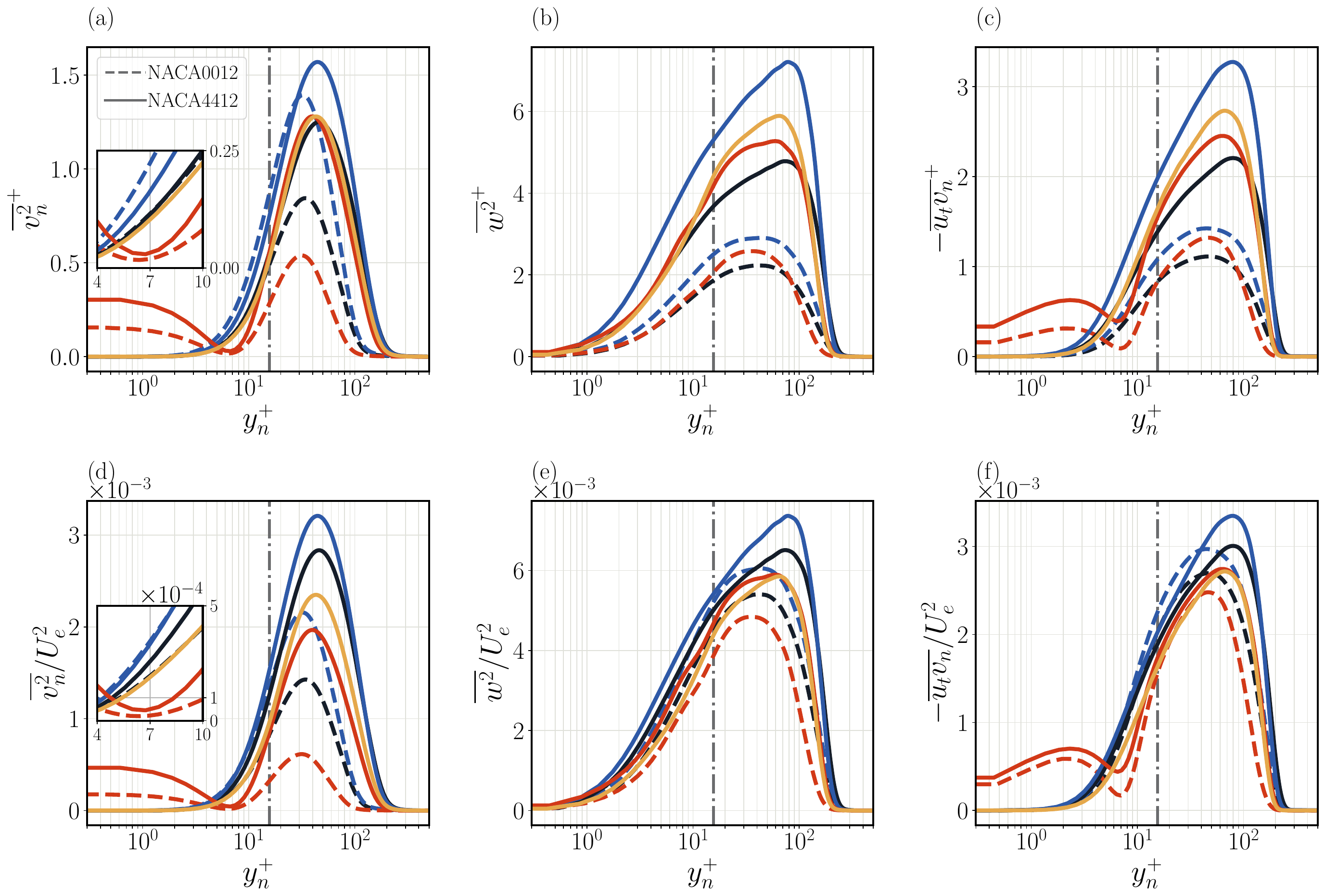}
    }
    \caption{(Top) Inner- and (bottom) outer-scaled fluctuations of (left) wall-normal velocity components 
            $\overline{v^2_n}$, (middle) spanwise velocity components  $\overline{w^2}$ and (right) Reynolds shear stress $\overline{u_t v_n}$
            as a function of the inner-scale wall-normal distance $y^+_{n}$ 
            at streamwise location of $x/c=0.75$  
            on suction side of NACA0012 (dashed lines) and NACA4412 (solid lines). 
            The dashed lines and solid lines denotes the configurations for NACA0012 and NACA4412, respectively. 
            The prescribed sensing plane of OC ($y^+_{s} = 15$) 
            is indicated by a gray dash-dotted line.          
            The color code follows table~\ref{tab:ctrl_configs}.
            }
    \label{fig:vv}
\end{figure}

Since OC is designed to mitigate convective momentum transport in the wall-normal direction~\citep{choi_active_1994}, 
it is highly relevant to examine wall-normal velocity fluctuations. 
The inner- and outer-scaled profiles at $x/c = 0.75$ are depicted in Figure~\ref{fig:vv}(a) and (d), respectively.

With increasing adverse pressure gradient, 
the inner-scaled wall-normal velocity fluctuations intensify, 
and the location {of the} maximum intensity moves {farther} from the wall. 
This position approximately aligns with the {outer-peak} location of $\overline{u^2_t}^+$~\citep{atzori_uniform_2021}. 
In particular, case {\rm Ref2} exhibits a more prominent outer peak, 
which is located approximately $10$ wall units farther from the wall compared to case {\rm Ref1}. 
The inherent {low-Reynolds-number} effect in case {\rm Ref1} does not significantly impact $\overline{v^2_t}$. 
Additionally, case {\rm Ref2} {exhibits} a higher outer peak in outer-scaled profiles 
due to intensified wall-normal convection caused by {the stronger APG}.

{For OC, the modifications of the outer peak {of the} inner-scaled profiles vary with the intensity of pressure gradient.  
In particular, the inner-scaled profile of case {\rm OC1} results in an intensified outer peak and case {\rm OC2} yields 
a slightly intensified outer peak of $\overline{v^2_n}^+$. 
However, OC significantly attenuates the outer peak of outer-scaled profiles, 
indicating that modifications of the outer peaks {of the} $\overline{v^2_n}^+$ are due to the variations in friction velocity.} 
Note that the outer peak exhibits an inward shift by approximately one wall unit, 
coinciding with the outward shift of $\overline{u^2_t}$. 
In the near-wall region, {the presence {of the} non-zero value at the wall is due to the control input~\citep{choi_active_1994}}, 
and further from the wall, 
the intensified $\overline{v^2_t}$ gradually decreases 
due to more intense momentum transport in the wall-normal direction. 
{In particular, examining the outer-scaled profiles at the wall, 
case {\rm OC2} yields $\overline{v^2_t}/U_e \approx 5\times10^{-4}$, 
which is more than twice {that of} case {\rm OC1} ($\approx 2\times10^{-4}$). 
The results indicate that the stronger {APG} induces higher control input.}

Furthermore, at $y^+_n \approx 7$, {the profiles} of OC {exhibit} local minima close to zero, 
as illustrated in the inset of each panel in figure~\ref{fig:vv}. 
This {wall-normal distance} corresponds to approximately half of the {wall-normal location of the} sensing plane $y^+_s = 15$, 
providing clear evidence of the so-called virtual wall~\citep{hammond_observed_1998}. 
\citet{hammond_observed_1998} outlined that the virtual wall results from 
the weakening of sweep and ejection events caused by opposition control, 
preventing convective momentum transport across the plane of the virtual wall. 
By prescribing the sensing plane at $y^+_s = 15$, 
\citet{hammond_observed_1998} reported the virtual wall in TCF at $y^+_s \approx 7.5$, 
and \citet{pamies_response_2007} found it at $y^+_s \approx 8.0$ in {the zero-pressure-gradient-turbulent} boundary layer, 
indicating good agreement with previous studies. 
The plane of the virtual wall shifts outward slightly with more intense {APGs}.

Uniform blowing intensifies wall-normal velocity fluctuations similarly to stronger APG, 
with both inner- and outer-scaled profiles showing increased and more prominent outer peaks, 
indicating enhanced wall-normal convection. 
{On the other hand}, body-force damping intensifies the outer peak of $\overline{v^2_n}^+$ 
due to reduced friction velocity. 
The outer-scaled profiles {result} in significantly attenuated outer peaks 
and reduced values at any wall-normal distance from the viscous sublayer.

The effects of pressure gradient and control schemes {on the} 
Reynolds shear stress $\overline{u_t v_n}^+$ and $\overline{u_t v_n}$ are similar to those {on the} wall-normal velocity fluctuations. 
Stronger adverse pressure {gradients intensify} the outer peak, moving it farther from the wall. 
Uniform blowing increases shear stress, equivalent to a stronger adverse pressure gradient. 
{Opposition control} and body-force damping attenuate Reynolds shear stress due to reduced $u_{\tau}$ below $y^+_n \approx 100$, 
but $|\overline{u_t v_n}|$ remains higher than the reference. 
The OC profiles also exhibit a local minimum at $y^+_n \approx 7$ with values close to zero, 
corresponding to the virtual wall position and linked to reduced wall-normal velocity fluctuations~\citep{hammond_observed_1998}. 
On the other hand, the spanwise velocity fluctuations $\overline{w^2}^+$ and $\overline{w^2}$ modifications 
caused by {APGs} and control schemes resemble those {on the} Reynolds shear stress. 
Note that OC does not produce a local minimum of $\overline{w^2}^+$ and $\overline{w^2}$ at the virtual wall plane.

\subsection{{Turbulent-kinetic-energy budgets}}
Following the analysis {of the} Reynolds-stress components, 
we investigate how the applied control schemes affect the {turbulent-kinetic-energy} (TKE) budget terms. 
Note that we {focus} on the TKE budget profiles at $x/c = 0.75$, 
which is a region characterized by a strong pressure gradient and minimal influence from moderate Reynolds numbers.

\begin{figure}
    \centering{
        \includegraphics[width=0.8\textwidth]{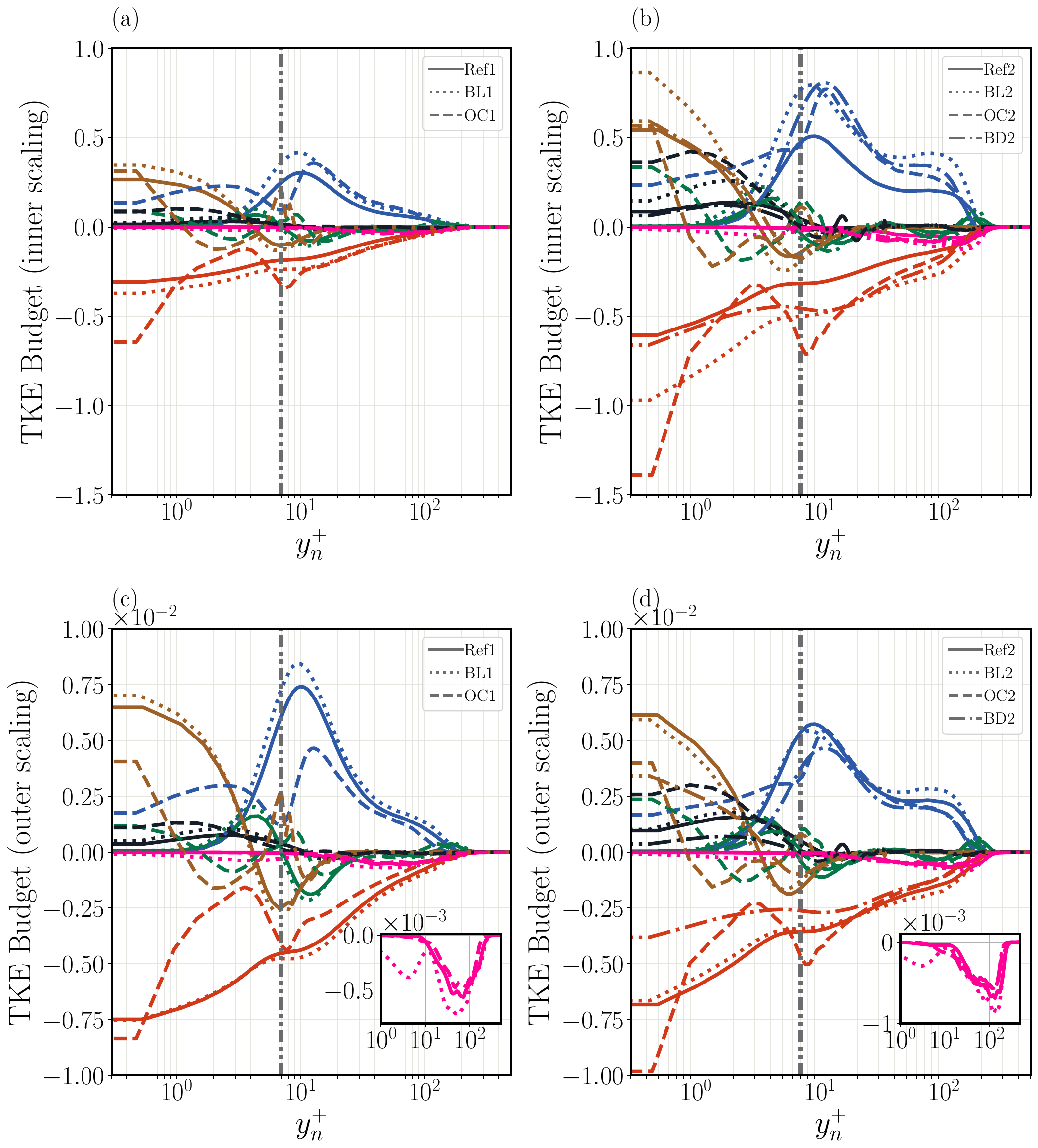}
    }
    \caption{
        (Top) Inner- and (bottom) outer-scaled {turbulent-kinetic-energy} budgets at $x/c = 0.75$.
        The panels {on the} left and right {correspond} to the suction side of NACA0012 and NACA4412, respectively. 
        The wall-normal location of $y^+_{n} = 7$ is indicated by dash-dotted {lines}. 
        The colours in the panels {correspond} to: 
        \textcolor{myblue}{production (blue)},
        \textcolor{myred}{dissipation (red)},
        \textcolor{mygreen}{turbulent transport (green)},
        \textcolor{mybrown}{viscous diffusion (brown)},
        \textcolor{myblack}{velocity-pressure-gradient correlation (black)} and 
        \textcolor{mypink}{convection (magenta)}.
            }
    \label{fig:TKE}
\end{figure}

The inner- and outer-scaled profiles of TKE budget terms at $x/c = 0.75$ 
on the suction side of NACA0012 and NACA4412 are depicted in figures~\ref{fig:TKE}(a,c) and (b,d), respectively. 
It can be observed that a stronger adverse pressure gradient significantly intensifies 
the inner-scaled profiles {of the} TKE budget terms. 
Specifically, {as the} APG increases, the inner-scaled profile of production exhibits 
a more pronounced inner peak and a local maximum in the outer region. 
This increase is related to the rise in turbulent shear stress in the outer region caused by APG~\citep{skaare_turbulentTKE_1994}, 
which is similar to the behavior of streamwise velocity fluctuations. 
To balance the intensified production, the inner-scaled profile of dissipation results in a significant outer peak.
The APG effect on large-scale motions in the outer layer 
leads to the redistribution of TKE terms near the wall~\citep{vinuesa_wingTBL_2017}. 
As the {pressure-gradient} intensifies, dissipation near the wall increases, 
while intensified viscous diffusion compensates for this. 
In the outer-scaled profiles, shown in figures~\ref{fig:TKE}(c) and (d), 
the shape of the profiles remains unchanged, 
but the magnitude decreases under stronger APG due to variations in friction velocity.

{Opposition control} significantly intensifies the inner-scaled profile of production at any wall-normal distance, 
{a fact which is consistent} with observations from the streamwise velocity fluctuations. 
The inner peak of production is intensified and shifts outward from the wall, 
while the outer peak increases due to {the reduced} $u_{\tau}$. 
This intensified production is compensated by stronger dissipation, leading {to an} amplified viscous diffusion.
At the wall, the control input results in non-zero production 
and a dramatic increase in dissipation, 
which is balanced by stronger viscous diffusion. 
The velocity-pressure-gradient correlation is also significantly enhanced, 
leading to a notable redistribution of TKE budget terms in the near-wall region. 
In particular, in case {\rm OC2}, the velocity-pressure-gradient correlation at the wall 
is twice {as large as that} of case {\rm OC1}, corresponding to the control input intensity observed {in the} wall-normal velocity fluctuations, 
which is linked to the fact that the reduction in friction is driven by the pressure gradient~\citep{choi_active_1994}.

Interestingly, at the virtual wall plane of $y^+ \approx 7$, 
the dissipation profile {exhibits} a small local {minimum}, 
and viscous diffusion {shows} a local maximum to balance the dissipation. 
This balance between dissipation and viscous diffusion at the virtual wall 
is similar to that observed at the wall, further indicating the formation of the virtual wall in the present study. 
Below the virtual wall, the {distributions} of viscous diffusion and turbulent transport {are} notably affected, 
showcasing negative values and local minima closer to the wall. 
Above the virtual wall, the {effect} of OC varies with pressure gradient intensity. 
Case {\rm OC1} achieves a more prominent reduction in the outer-scaled production term 
compared to the reference at any wall-normal distance greater than $y^+_n \approx 7$, 
{which is linked to the} energized large-scale motion in the outer region caused by stronger {APGs}. 
Since OC is designed to mitigate velocity fluctuations near the wall, 
it is less effective in counteracting intensified wall-normal convection caused by stronger APG, 
making large-scale motion in the outer region difficult to suppress under strong APG conditions.

It is worth noting that we also {assessed the effect of choosing different sensing planes throughout the boundary layer, 
including in the outer region, on the performance of the OC. 
Our results (not shown here) showed that,} as suggested by~\citet{hammond_observed_1998}, 
prescribing a sensing plane too far from the wall leads to instability of control.
In particular, OC results in increased friction drag, which is more pronounced as the sensing plane {gets farther from the wall}.

The effects of body-force damping on TKE budgets are simpler than those of OC, 
as there is no virtual wall {in this control technique}. 
In particular, {the inner-scaled} profiles of TKE budgets are intensified due {to the} reduced friction velocity, 
{with the} outer-scaled profiles exhibiting an outward shift by approximately one wall unit. 
On the contrary, uniform blowing further intensifies the effects of a strong adverse pressure gradient.

Furthermore, the outer-scaled mean convection profiles shown in figures~\ref{fig:TKE}(c) and~\ref{fig:TKE}(d) reveal that 
uniform blowing intensifies the convection term due {to the increased} wall-normal {velocity,} 
while OC and body-force damping do not significantly modify convection profiles. 
These results are consistent {with the previous} observations on mean and fluctuating components {of the} wall-normal velocity.

\section{Spectra analysis}\label{sec:spectra}
So far, the assessment of the control mechanisms indicates that adverse pressure gradient and opposition control 
have opposite impacts {on the} wall-normal momentum transport, 
{with the} APG intensifying it while OC is designed to mitigate it. 
Therefore, it is of great interest to investigate this interaction 
in terms of the energy distribution {across} the scales via spectral analysis.

In the present study, time series of the velocity components were obtained for a number of wall-normal profiles{, spanning} a total of $10$ flow-over times 
with a sampling rate of $10^{-3}$ and $7.5 \times 10^{-4}$ flow-over times {for the} NACA0012 and NACA4412 cases, respectively. 
We focus on the $x/c= 0.75$, 
and report the one- and two- dimensional power-spectral densities, 
which are computed {using the} fast Fourier transform (FFT). 

\subsection{One-dimensional spanwise power-spectral densities}
\begin{figure}
    \centering{
        \includegraphics[width=\textwidth]{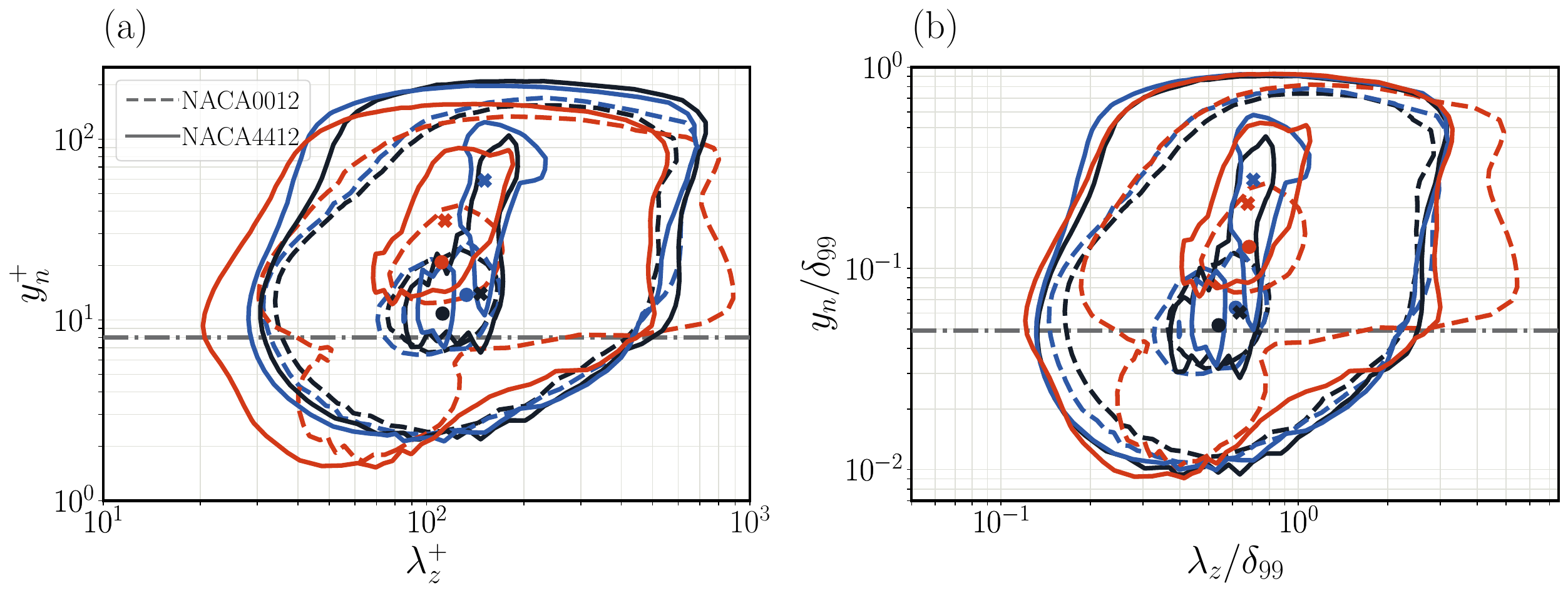}
    }
    \caption{Inner-scaled premultiplied spanwise power-spectral density (PSD) of the wall-tangential {velocity fluctuation, in terms of} 
            (left) the inner-scaled spanwise wavelength (${\lambda}^+_{z}$) and wall-normal distance ($y^+_n$), 
            and (right) of the outer-scaled spanwise wavelength ($\lambda_{z} / \delta_{99}$) and the wall-normal distance ($y_{n}/\delta_{99}$){. Results shown} 
            at $x/c = 0.75$ on the suction side {of the} NACA0012 (dashed lines) and NACA4412 (solid lines) {cases}. 
            The contours illustrate the levels corresponding to $15\%$ and $75\%$ of the maximum power density in the inner region for each case, 
            and the locations of the maxima achieved {on the} NACA0012 and NACA4412 are marked with circles and crosses, respectively. 
            The color code follows table~\ref{tab:ctrl_configs}. 
            }
    \label{fig:sp1D}
\end{figure}

Figure~\ref{fig:sp1D}(a) and (b) depict the contours of the inner-scaled premultiplied {power-spectral} density of 
wall-tangential velocity fluctuations ($k_{z} \phi^+_{u_t u_t}$) at $x/c = 0.75$, 
expressed as a function of spanwise wavelength $\lambda{z}$ and wall-normal distance $y_n$ in inner and outer units, respectively.

Under stronger {APGs}, higher small-scale energy is produced in the near-wall region, 
and {also more} large-scale energy is observed in the outer region, {leading to a more prominent spectral} outer peak. 
This behavior, reported by~\citet{tanarro_effect_2020} {using a} similar analysis at a higher ${Re}_c = 400 \ 000$, 
is also evident in our results when comparing the spectra of {\rm Ref1} and {\rm Ref2}.

{Opposition control} significantly alters both the value and location of the spectral inner peak. 
In particular, case {\rm Ref1} exhibits {a magnitude of the} inner peak of $4.68$, 
whereas case {\rm OC1} results in a lower inner peak of $4.22$, 
located at a higher wall-normal distance and slightly lower inner-scaled wavelength $\lambda^+_z$. 
Note that there is no outer peak for case {\rm Ref1} due to the lower pressure gradient. 
In contrast, case {\rm Ref2} exhibits inner and outer peaks of $5.89$ and $4.82$, respectively, 
while case {\rm OC2} enhances these values to $5.93$ and $5.37$, 
which are located {farther} from the wall and at lower $\lambda^+_z$. 
Additionally, OC attenuates energy in the outer region for any inner-scaled wavelength.

In the near-wall region, the effects of OC {on the power-spectral} density 
vary significantly with the adverse pressure gradient. 
For case {\rm OC1}, below the virtual wall, only small-scale energy (ranging from $50 \leq \lambda^+_z \leq 100$) {is present}, 
indicating that OC effectively suppresses wall-normal transport of streamwise velocity fluctuations under mild APG, 
but further mitigation of small-scale structures remains challenging. 
Above the virtual wall, case {\rm OC1} significantly energizes {the large} scales in the near-wall region 
and also energizes small scales, {although} less intensely. 
On the other hand, case {\rm OC2} {exhibits} a wider range of scales below the virtual wall, 
with less pronounced large-scale energy but further energized small scales.
Above the virtual wall, case {\rm OC2} further energizes small-scale structures below $\lambda^+_z < 30$ 
without significantly modifying energy for scales larger than $\lambda^+_z \approx 400$ in the near-wall region.

When {the power-spectral} densities are {represented} as functions of outer-scaled wavelength and wall-normal distance, 
the energization {of the} small-scale structures is less pronounced. 
Below the virtual wall, case {\rm OC2} decreases energy for scales larger than $\lambda_z / \delta_{99} = 0.35$ 
and smaller than $0.65$, 
whereas case {\rm OC1} shows energy only for $0.45 \leq \lambda_z / \delta_{99} \leq 0.65$. 
This indicates that OC may not effectively suppress the enhanced wall-normal transport of streamwise velocity fluctuations 
caused by stronger APG due to intensified transport of small-scale structures. 
Additionally, energy attenuation in the outer region varies with outer-scaled wavelength and wall-normal distance. 
Specifically, case {\rm OC1} increases energy in the outer region, 
while the {power-spectral} density of case {\rm OC2} aligns with case {\rm Ref2}.

The effects of uniform blowing have been extensively studied by~\citet{atzori_uniform_2021}, 
who reported that uniform blowing increases {power-spectral} density, 
with stronger effects in the outer region, resulting in the {largest} outer peak. 
Similar observations are made in the present data set, 
{{although} the different intensity of pressure gradient leads to distinct performance.} 
{For case {\rm BL1}, the inner peak {has a} magnitude of $5.25$ with a {larger} inner-scaled wavelength than the uncontrolled case. }
In case {\rm BL2}, {both the} inner and outer peaks are {higher, with magnitude of} $8.09$ and $7.94$, respectively, 
with higher wall-normal distances, but slightly lower $\lambda^+_z$ for both peaks.

\begin{figure}
    \centering{
        \includegraphics[width=0.95\textwidth]{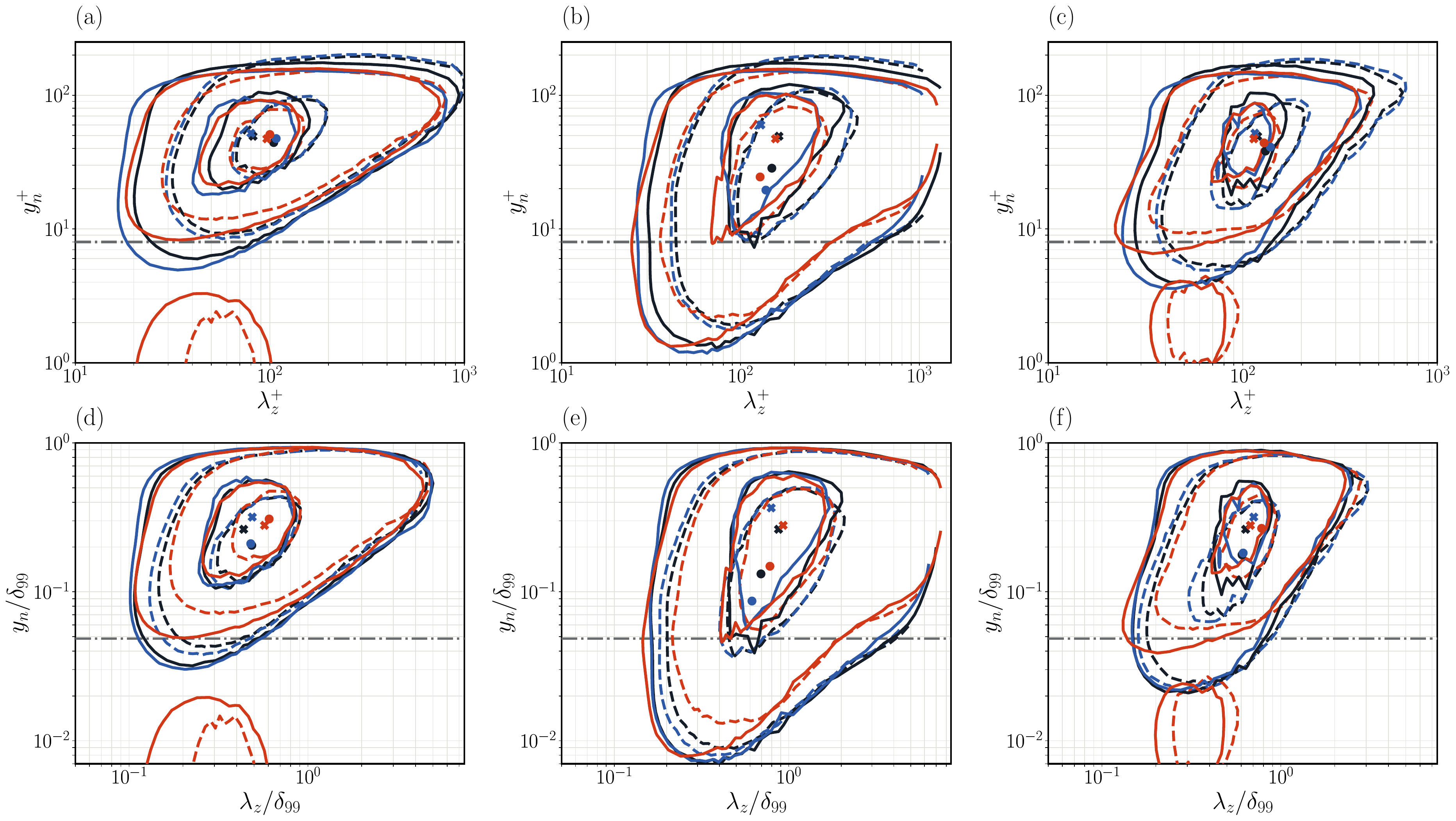}
    }
    \caption{Inner-scaled premultiplied spanwise power-spectral density (PSD) of 
            the (left) wall-normal, (middle) spanwise and (right) Reynolds-shear stress
            as functions of (top) the inner-scaled spanwise wavelength (${\lambda}^+_{z}$) and wall-normal distance ($y^+_n$), 
            and (bottom) of the outer-scaled spanwise wavelength ($\lambda_{z} / \delta_{99}$) and the wall-normal distance ($y_{n}/\delta_{99}$)  
            for all the considered case 
            at $x/c = 0.75$ on the suction side of NACA0012 (dashed lines) and NACA4412 (solid lines). 
            The contours illustrate the levels corresponding to the $20\%$ and $80\%$ of the maximum {power-spectral} density in the inner region for each case, 
            and the locations of the maxima achieved on NACA0012 and NACA4412 are marked with circles and crosses, respectively. 
            The color code follows table~\ref{tab:ctrl_configs}. 
            }
    \label{fig:sp1D_vv_uv_ww}
\end{figure}

Furthermore, we assess the {power-spectral} densities of {$k_{z} \phi^+_{{v_{n} v_{n}}}$, $k_{z}\phi^+_{{w w}}$, and $k_{z}\phi^+_{{u_{t} v_{n}}}$} 
at the same streamwise location in figure~\ref{fig:sp1D_vv_uv_ww}. 
These are expressed as functions of spanwise wavelength $\lambda_{z}$ and wall-normal distance $y_n$.

The adverse pressure gradient generally {energizes the} small scales in the near-wall region 
and intensifies the convection of small-scale structures farther from the wall. 
Uniform blowing further energizes the small scales near the wall, 
intensifies wall-normal convection, 
and results in even {larger power-spectral-density} values.

The effects of OC vary with the intensity of the pressure gradient. 
Above the virtual wall, case {\rm OC1} generally energizes small scales and attenuates large scales near the wall, 
corresponding to the attenuation of all scales when the {power-spectral} densities 
are shown as functions of the outer-scaled wavelength and wall-normal distance. 
However, case {\rm OC2} not only attenuates the large scales 
but also further energizes the small scales in the near-wall region above the virtual wall, 
being consistent with observations of the spectra as functions of the outer-scaled wavelength and wall-normal distance.

Below the virtual wall, {$k_{z}\phi^+_{{v_n v_n}}$} under opposition control shows energized small scales, 
corresponding to the control input at the wall and mitigated momentum transport in the wall-normal direction. 
In particular, case {\rm OC2} results in more energetic small scales ranging from {$20 \leq \lambda^+_{z} \leq 100$} up to {$y^+_n \approx 3$} 
compared to case {\rm OC1}, which ranges from {$40 \leq \lambda^+_{z} \leq 90$} up to $y^+_n \approx 2.5$. 
Interestingly, this wavelength range roughly corresponds to that of the reference below the virtual wall, 
indicating that OC effectively suppresses wall-normal transport of {the wall-normal fluctuations}. 
Our results also suggest that smaller near-wall structures caused by stronger APG induce the intensification of control input at the wall.

Similar observations can be made for {$k_{z}\phi^+_{{u_t v_n}}$} below the virtual wall, 
where OC results in energized small-scale structures. 
In particular, the energized spectra of cases {\rm OC1} and {\rm OC2} 
range approximately from {$37 \leq \lambda^+_{z} \leq 70$} and {$45 \leq \lambda^+_{z} \leq 100$} up to {$y^+_n \approx 3$}, respectively. 
This is linked to the combined effects of opposition control on {$k{z}\phi^+_{{u_t u_t}}$} and {$k_{z}\phi^+_{{v_n v_n}}$}.

\subsection{Two-dimensional power-spectral densities}
\begin{figure}
    \centering{
        \includegraphics[width=\textwidth]{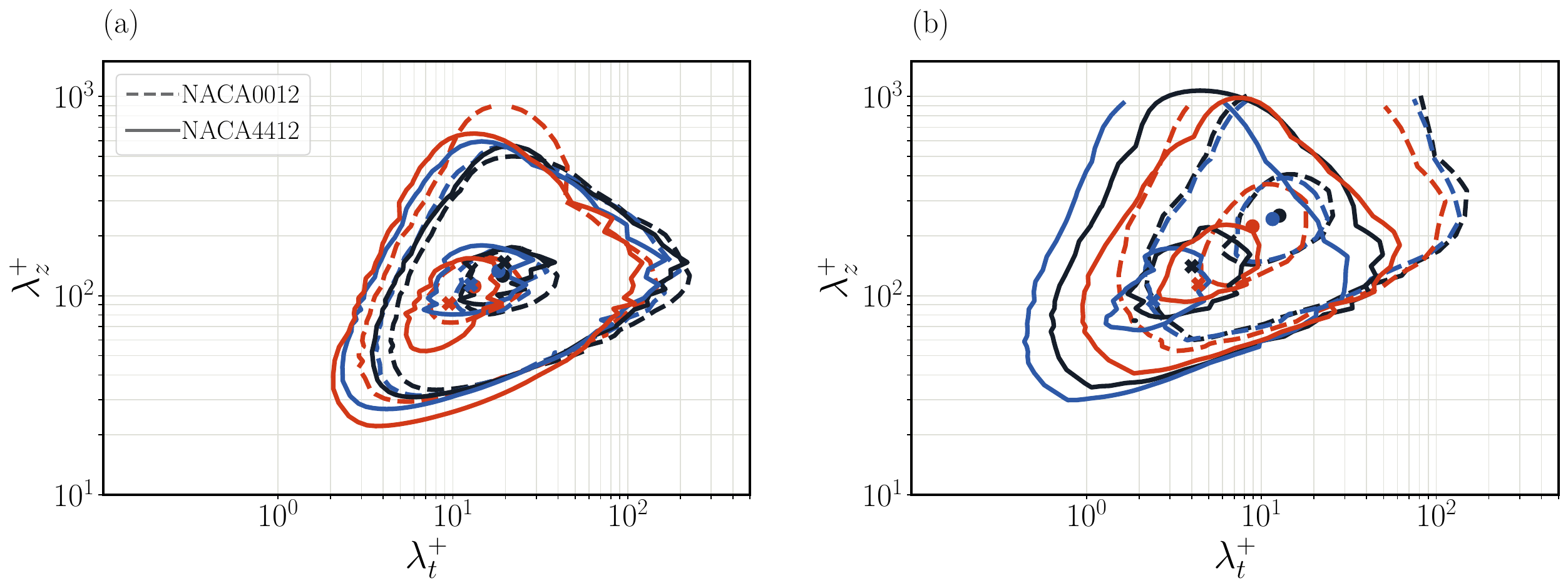}
    }
    \caption{
    Inner-scaled premultiplied spanwise and temporal power-spectral density (PSD) of the wall-tangential velocity fluctuations 
    at (left) $y^+_n = 15$ and (right) $y^+_n = 150$
    for all the considered {cases} 
    at $x/c = 0.75$ on the suction side of NACA0012 (dashed lines) and NACA4412 (solid lines). 
    The contours illustrate the levels corresponding to $15\%$ and $75\%$ of the maximum power density in the inner region for each case, 
    and the locations of the maxima achieved on {the} NACA0012 and NACA4412 are marked with circles and crosses, respectively. 
    The gray dashed-dot lines indicate the wall-normal position corresponding to $y^+_n = 7$. 
    The color code follows table~\ref{tab:ctrl_configs}.
    }
    \label{fig:sp2D}
\end{figure}

In this section, we assess the two-dimensional {power-spectral} densities comprising spanwise and temporal information from the time series.
Figure~\ref{fig:sp2D} depicts the contours of the two-dimensional spectral densities of 
the streamwise velocity fluctuations for opposition control, uniform blowing, and uncontrolled cases 
on the suction side at $x/c = 0.75$ and at two wall-normal distances: {$y^+_{n} = 15$} and {$y^+_{n} = 150$}. 
Note that we employ Welch's overlapping window method for computing the {power-spectral} density due to the lack of periodicity in the {temporal data}.

At $y^+_n = 15$, the intensified APG modifies both the intensities and shapes of the two-dimensional spectra. 
In particular, case {\rm Ref2} results in higher spectral density values of approximately $2.6$ compared to $2.0$ for case {\rm Ref1}. 
Additionally, $\lambda^+_t$ of {\rm Ref2} is reduced, 
and the smallest and largest structures are expanded towards both shorter and longer wavelengths, respectively. 
{Opposition control} reduces the {power-spectral} density values to approximately $1.8$ and $2.1$ for cases {\rm OC1} and {\rm OC2}, respectively, 
while the spectra are located at lower $\lambda^+_z$ and $\lambda^+_t$. 
Note that case {\rm OC1} further energizes large-scale structures towards longer wavelengths and periods. 
Uniform blowing amplifies the {power-spectral} density values, 
with cases {\rm BL1} and {\rm BL2} reaching the highest values of $2.3$ and $2.9$, respectively. 
In case {\rm BL1}, the location is not drastically modified, but in case {\rm BL2}, it is shifted to lower $\lambda^+_z$ and $\lambda^+_t$.

At $y^+_n = 150$, 
the stronger APG {leads to an extension of the lower-energy} contour towards smaller wavelengths and shorter periods. 
This is linked to the fact that {a} stronger APG intensifies wall-normal convection and transports small-scale structures to the outer region. 
Note that case {\rm Ref2} does not energize large structures with high periods, 
differing from the observations by~\citet{tanarro_effect_2020} at ${Re}_c = 400 \ 000$, 
where {the} stronger APG energized all scales. 
In the present study, this may be due to the lower Reynolds number, leading to less separation of scales.

Opposition control again reduces the {power-spectral} density values, 
but its effects on the contour shape vary under different {pressure gradients}. 
In particular, case {\rm OC1} increases the smallest structures, 
showcasing a similar effect to a stronger adverse {pressure gradient}, 
while {the energy contained within the} case {\rm OC2} exhibits the opposite effect. 
Uniform blowing increases the {power-spectral} density values. 
However, in case {\rm BL1}, the spectra are drastically modified by the control, 
with all structures slightly attenuated. 
The contour of case {\rm BL2} is similar to that of case {\rm OC2}, 
but with higher energization of small-scale structures with low periods.

\section{Summary and conclusions}\label{sec:conclusions}
In the current study, we analyzed the effects of opposition control 
applied to the spatially developing turbulent boundary layer (TBL) 
over the suction side of the NACA0012 wing section at $0^{\circ}$ angle of attack and the NACA4412 wing section at $5^{\circ}$ angle of attack, 
both at ${Re}_c = 200 \ 000$. 
The objective was to understand how the control scheme interacts with mild and strong adverse pressure gradients, respectively.

Overall, the performance of opposition control is significantly influenced {by the} adverse pressure gradient: 
opposition control is designed to mitigate near-wall fluctuations, 
but adverse pressure gradient further amplifies them by intensifying wall-normal convection. 
Additionally, uniform blowing was considered on both wing sections, 
and {a database} {where} body-force damping {is applied on} the NACA4412 wing section~\citep{atzori_uniform_2021} 
was used for comparison among different {active-flow-control} schemes aimed at reducing friction drag. 
Despite differences in {friction-drag-reduction} performance, 
the effects of uniform blowing are similar to those of a stronger APG, 
while the effects of body-force damping are more similar to those of opposition control 
in terms of integral quantities rather than turbulence statistics.

The Clauser pressure-gradient coefficient $\beta$ quantitatively indicates that 
the TBL over the suction side of the NACA4412 is subjected to a much stronger APG than that of the NACA0012. 
This results in significant differences in the streamwise development {of the} various integral quantities. 
Opposition control reduces both boundary-layer thickness and friction, 
leading to higher $\beta$ and $H$ as well as lower $Re_{\theta}$, 
which are more pronounced {for the} TBL subjected to milder pressure gradient. 
On the other hand, the achieved reduction in friction drag also varies with the intensity of the pressure gradient. 
Despite the maximum relative reduction of the skin-friction coefficient ($R$) 
obtained at the edge of the control section due to the {change of wall} boundary condition, 
opposition control achieves maxima of $R \approx 40\%$ and $\approx 30\%$ on the NACA0012 and NACA4412, respectively. 
{As} the pressure gradient {gets} more intense, the $R$ achieved {by the} opposition {control gets smaller}.
{As the} opposition control {gets} a lower drag coefficient $C_{d}$ by reducing both $C_{d,f}$ and $C_{d,p}$, 
{also leading to a} slightly higher $C_{l}$, {it leads to an improved} aerodynamic efficiency ($L/D$). 
Specifically, opposition control reduces $C_{d}$ by $6.8\%$ and $5.1\%$ 
compared to the reference $C_d$ of NACA0012 and NACA4412, respectively.

The effects of opposition control are further investigated 
by analyzing the inner- and outer-scaled wall-normal turbulence statistics. 
First, we assess the mean wall-tangential velocity $U_t$ at $x/c = 0.4$ and $0.75$, 
where the stronger APG energizes the wake, 
resulting in higher $U_{t}$ in the outer region. 
The manipulated inner-scaled mean wall-tangential velocity results in a more prominent outer region, 
while in the viscous sublayer, the controlled profiles of $U_t$ are vertically shifted downwards with respect to the reference. 
Additionally, the comparison between inner- and outer-scaled wall-normal mean velocity ($V_n$) profiles at $x/c = 0.75$ 
reveals that opposition control does not drastically affect mean wall-normal convection.

Subsequently, we assess the Reynolds-stress terms at $x/c = 0.75$. 
Generally, the inner-scaled profiles of opposition control result in intensified inner peaks, 
which are slightly shifted outwards for streamwise velocity fluctuations and inwards {for the} other terms. 
This is less pronounced when the pressure gradient intensifies. 
Opposition control also energizes the outer peak of inner-scaled streamwise velocity fluctuations 
caused by a stronger pressure gradient, which is independent of the scaling.

One critical insight is the presence of the so-called virtual wall~\citep{hammond_observed_1998}, 
manifested by examining the local minima of the wall-normal velocity fluctuation $\overline{v^2_{n}}$ profiles in the viscous sublayer. 
In the present study, the plane of the virtual wall is at a wall-normal distance of $y^{+}_{n} = 7$, 
being roughly half of the prescribed sensing plane ($y^{+}_{n} = 15$), 
which reflects the effective mitigation of wall-normal momentum transport achieved by opposition control. 
The presence of the virtual wall significantly modifies 
the profile shapes as a consequence of mitigating wall-normal momentum transport. 
Note that the virtual wall's presence is also reflected in the $\overline{u^2_t}$ and $\overline{u_t v_n}$ profiles.

The analysis of the TKE budgets shows that opposition control leads to a significant redistribution of TKE budget terms. 
The outer-scaled profiles of the production term for opposition control exhibit a remarkable decrease in the inner peak above the virtual wall, 
which is more prominent for the controlled TBL subjected to milder pressure gradient. 
Opposition control also suppresses the outer peak in the outer-scaled production profile, but with relatively less intensity. 
This suggests that the amplified production is strongly linked to the downgraded control performance under a stronger pressure gradient.
Furthermore, opposition control balances dissipation and viscous diffusion at the plane of the virtual wall, 
similar to observations at the physical wall, 
further implying the formation of the virtual wall~\citep{hammond_observed_1998,ge_dynamicTCF_2017}.

The results of spectra analysis are discussed as an extension of the previous investigation on {turbulence} statistics.
First, we assess the one-dimensional {power-spectral} densities of inner-scaled Reynolds-stress tensor terms 
as a function {of the} spanwise wavelength $\lambda_z$. 
In general, opposition control results in lower {power-spectral} densities 
and attenuates large-scale structures below the plane of the virtual wall. 
However, due to the transport of small scales from the near-wall region to the outer layer 
induced by APG~\citep{tanarro_effect_2020}, 
the interaction between opposition control and APG 
in terms of the energization of structures significantly varies with the intensity of the pressure gradient, 
as reflected by examining the spectral contours with lower $\lambda_z$. 
Essentially, when the manipulated TBL is subjected to milder APG, 
opposition control attenuates a wider range of wavelengths in the near-wall region. 
The structures in the outer region are not drastically modified by opposition control.
Subsequently, we assess the two-dimensional spectra of 
inner-scaled streamwise velocity fluctuations 
as a function of spanwise wavelength and temporal information at $y^+_n = 15$ and $150$. 
The results further confirm that the spectral contour of opposition control is 
significantly affected by the energization of small scales in the outer region caused by APG, 
which is more prominent for shorter periods.


One remaining open question is how to improve the performance of opposition control applied to APG TBLs. 
In principle, higher drag reduction might be achieved through a combined-control method 
that imposes fluctuations of wall-normal and spanwise velocity components at the wall, 
as reported by~\citet{choi_active_1994,wang_active_2016}. 
However, since opposition control becomes unstable when the prescribed sensing plane is too far from the wall~\citep{choi_active_1994,hammond_observed_1998}, 
it remains challenging for effectively mitigating the energized outer region by suppressing wall-normal momentum transport. 

The current study describes the interaction of opposition control and TBLs subjected to nonuniform adverse pressure gradient. 
To the best of our knowledge, this is the first study to apply opposition control to {turbulent wings} 
and {to conduct} an in-depth analysis of its effects on such complex geometries.

\backsection[Acknowledgements]{The simulations were run with the computational resources provided by the Swedish
National Infrastructure for Computing (SNIC). 
{The authors thank Raffaello Mariani for his insightful comments on this manuscript.}
}

\backsection[Funding]{This work is supported by the funding provided by the Swedish e-Science Research Centre (SeRC),
ERC grant no. `2021-CoG-101043998, DEEPCONTROL' to RV. 
{The views and opinions expressed are however those of the author(s) only and do not necessarily reflect those of European Union or European Research Council.} }

\backsection[Declaration of interests]{The authors report no conflict of interest.}

\backsection[Data availability statement]{The data that support the findings of this study are openly available on GitHub-\href{https://github.com/KTH-FlowAI}{KTH-FlowAI}}

\backsection[Author ORCIDs]{\\
{Yuning Wang}       \href{https://orcid.org/0009-0009-9964-7595}{https://orcid.org/0009-0009-9964-7595};\\
{Marco Atzori}      \href{https://orcid.org/0000-0003-0790-8460}{https://orcid.org/0000-0003-0790-8460};\\
{Ricardo Vinuesa}   \href{https://orcid.org/0000-0003-0820-7009}{https://orcid.org/0000-0003-0820-7009}.
}

\appendix
\section{Skin-friction contributions}\label{sec:FIK}
In this section, 
to further investigate the contributions of opposition control to the skin friction in the TBL 
{by means of the so-called} FIK identity proposed by~\citet{fukagata_lower_2009}.
In the light of the previous work by~\citet{atzori_new_2023}, 
we employ both the ``standard'' formulation, 
derived from the conservative form of the averaged {Navier--Stokes} equations, 
and the ``boundary-layer'' formulation, 
which offers a new perspective on evaluating the contributions of the pressure gradient.
Detailed derivations of these formulations can be found 
in the {studies by}~\citep{renard_FIKBL_2016,atzori_new_2023} and will not be presented here. 
Additionally, we incorporate results from previous 
investigations of uniform blowing and body-force damping 
on the suction side of the NACA4412 wing section for comparison.

\subsection{Standard formulation}
The terms {of the} standard formulation are obtained 
by integrating {by parts of the} momentum equation on the wall-normal profiles up to $y_n = \delta_{99}$. 
The decomposition of skin friction $c_f$ results in four components~\citep{renard_FIKBL_2016}, which are given by
{\begin{eqnarray}
        c_f (x_t) &=&  \underbrace{\frac{4 (1 - {\delta}^* / {\delta}_{99})}{{Re}_{\delta}}}_{c^{\delta}_{f} (x_t)} \
                    \underbrace{-4 \int^1_0 (1 - \eta) \frac{\overline{u_t v_n}}{U^2_e} {\rm d} \eta}_{c^{T}_{f} (x_t)}  \ 
                  \underbrace{-2 \int^1_0 {(1 - \eta)}^2 \frac{\delta_{99}}{U^2_e} \frac{\partial P}{\partial x_t} {\rm d}\eta}_{c^{P}_{f} (x_t)} \\    
                  &&  \underbrace{-2 \int^1_0 {(1 - \eta)}^2 \frac{\delta_{99}}{U^2_e} \left ( \frac{\partial U_t^2}{\partial x_t} + 
                  \frac{\partial (\overline{u^2_t})}{\partial x_t} + 
                  \frac{\partial U_t V_n}{\partial y_n} - \frac{1}{{Re}_{\delta}}\frac{\partial^2 U_t}{\partial x^2_t} \right)  {\rm d} \eta}_{c^{D}_{f} (x_t)}, 
    \label{eq:FIK_standard}
\end{eqnarray}}
\noindent where the integration variable $\eta = y_n / \delta_{99}$ is the wall-normal distance normalized with {the $99\%$} boundary-layer thickness. 

The first contribution $c^{\delta}_f$ is strongly linked to the evolution of boundary-layer thickness. 
The second term $c^{T}_f$ takes the Reynolds shear stress into account. 
The third contribution $c^{P}_f$ explicitly describes the effect {of the} pressure gradient. 
And the forth term $c^D_f$ includes all the terms in the streamwise momentum equation, 
which should equal to zero if the flow is homogeneous in the streamwise direction. 

\begin{figure}
    \centering{
        \includegraphics[width=\textwidth]{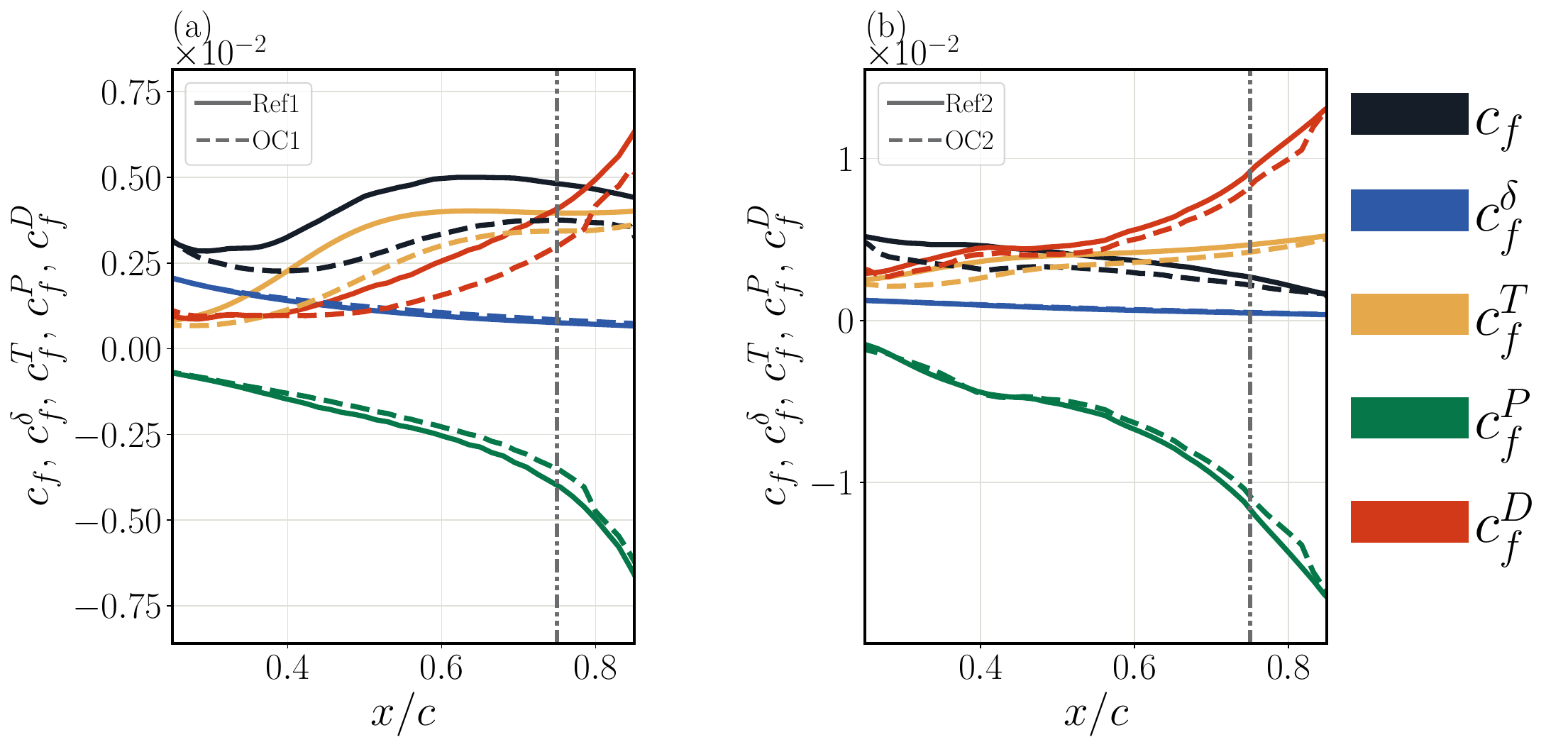}
    }
    \caption{Streamwise development of the componental contribution to the skin friction within the control section 
            on the suction side of NACA0012 (a) and NACA4412 (b). 
            The uncontrolled and OC profiles are denoted by solid and dashed lines, respectively.
            The dash-dotted line indicates the streamwise location of $x/c=0.75$, where the relative change of components are reported in table~\ref{tab:FIK_075}.
            The color code of each components are reported at the right of the panel (b). 
        }
    \label{fig:standard_FIK}
\end{figure}

Figure~\ref{fig:standard_FIK}(a) and~\ref{fig:standard_FIK}(b) depict the streamwise evolution of 
the four components on the suction side of NACA0012 and NACA4412, respectively. 
In particular, the $c^{\delta}_f$ exhibits small but non-trivial contribution due to the moderate Reynolds number in the present study, 
whereas {the control effect} is less pronounced on this term.  

The Reynolds shear stress contribution ($c^T_f$), 
exhibits a similar order of magnitude to the total $c_f$. 
On the suction side of the NACA0012, 
all $c^T_f$ profiles show a similar trend of streamwise evolution as the total $c_f$, 
especially after $x/c = 0.4$. 
However, on the suction side of the NACA4412, 
$c^T_f$ increases monotonically due to the strong APG enhancing the turbulent intensity as the boundary layer develops. 
OC significantly reduces $c^T_f$ as it is designed to suppress the turbulence fluctuations in the vincinity of the wall.

The pressure-gradient term $c^P_f$ is the only component that 
yields a negative contribution to the total skin friction, 
attributed to the positive pressure gradient {${\rm d} P/{\rm d} x_t$} on the suction side. 
This negative contribution intensifies with a stronger pressure gradient, 
contributing to the lower total $c_f$ for NACA4412 after $x/c = 0.5$ (see Figure~\ref{fig:Cf_and_R_and_E}(a)). 
{Note that OC} suppresses this contribution. 
As this term is independent {of the} wall-shear stress, 
the results suggest that the increase in the Clauser pressure-gradient {parameter} ($\beta$) 
{induced} by OC is mainly due to the reduction of $\tau_w$.

\begin{figure}
    \centering{
        \includegraphics[width=\textwidth]{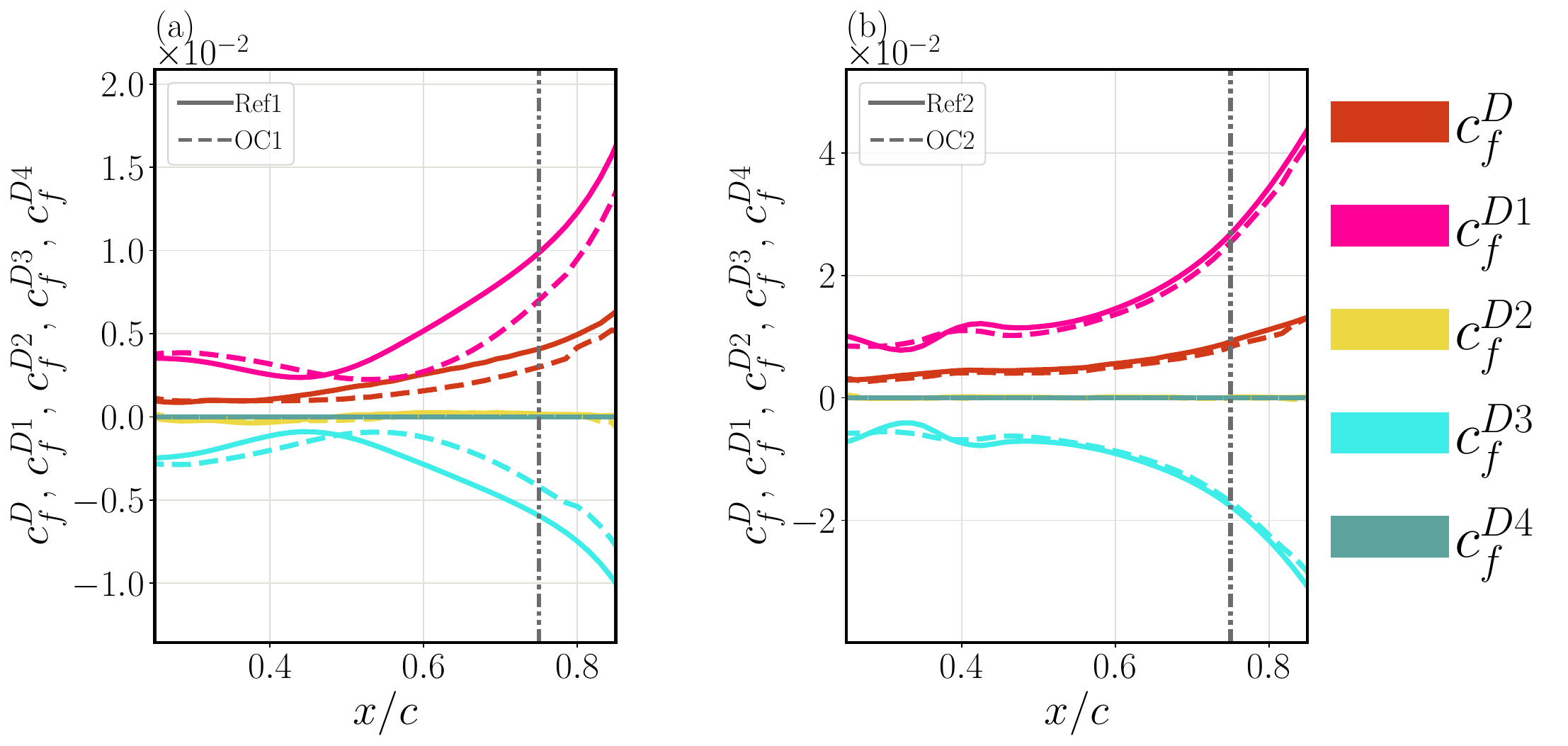}
    }
    \caption{Streamwise development of the decomposition of the $C^D_f$ within the control section on the suction side of NACA0012 (a) and NACA4412 (b). 
    The uncontrolled and OC profiles are denoted by solid and dashed lines, respectively.
    The dash-dotted line indicates $x/c=0.75$, where the relative change of components {is} reported in table~\ref{tab:FIK_075}.
    The color code of {each of the} components {is} reported at the right of the panel (b).}
    \label{fig:CD_1234}
\end{figure}

The contribution of streamwise homogeneity, $C^D_f$, 
exhibits significant growth towards the trailing edge due to the effect of the pressure gradient~\citep{atzori_uniform_2021}, 
eventually exceeding the total $c_f$ at the end of the control region. 
On the suction side of the uncontrolled NACA4412, 
$C^D_f$ exhibits more prominent growth than on the NACA0012, 
indicating that the evolution of $C^D_f$ is closely linked to the intensity of the pressure gradient. 
Although the applied control schemes effectively reduce this contribution, 
further investigation is {required} to understand how the control mechanisms affect different terms in $C^D_f$.
According to {equation~(\ref{eq:FIK_standard})}, the $C^D_f$ can be further split into four terms expressed as: 
{\begin{eqnarray}
    {c^{D1}_f}  &=&{-2 \int^1_0 {(1 - \eta)}^2 \frac{\partial U_t U_t}{\partial x_t} {\rm d} \eta }, \\ 
    {c^{D2}_f}  &=&{-2 \int^1_0 {(1 - \eta)}^2 \frac{\partial \overline{u_t^2}}{\partial x_t} {\rm d} \eta }, \\ 
    {c^{D3}_f}  &=&{-2 \int^1_0 {(1 - \eta)}^2 \frac{\partial U_t V_n}{\partial \eta} {\rm d} \eta }, \\
    {c^{D4}_f}  &=&{-2 \int^1_0 {(1 - \eta)}^2  \left (- \frac{1}{{Re}_{\delta}}\frac{\partial^2 U}{\partial x^2_t} \right ) {\rm d} \eta }, 
    \label{eq:FIK_CD}
\end{eqnarray}}
\noindent such that $c^{D}_f = c^{D1}_f + c^{D2}_f + c^{D3}_f + c^{D4}_f$.
The four components of $C^D_f$ are reported in Figure~\ref{fig:CD_1234}(a) and~\ref{fig:CD_1234}(b) 
for the suction side of NACA0012 and NACA4412, respectively. 

In particular, the contributions of $c^{D2}_f$ and $c^{D4}_f$ are negligible 
because the value of the second derivative of wall-tangential fluctuations is relatively low 
and $c^{D4}f$ is proportional to the inverse of ${Re}_{\delta}$. 
Therefore, due to the presence of the mean velocity components {of the} 
$c^{D}_f$ is dominated by $c^{D1}_f$ and $c^{D3}_f$, 
which correspond to the streamwise development of the streamwise mean velocity and wall-normal convection, respectively. 
The $c^{D1}_f$ contribution exhibits a higher magnitude than $c^D_f$ 
and is positive due to the negative {term} {${\rm d}{U_t}/{\rm d}{x_t}$}. 
It rapidly increases towards the trailing edge as the APG becomes stronger, following a similar trend to $c^D_f$. 
On the other hand, the strong negative $c^{D3}_f$ balances the spatial development of $c^{D1}_f$. 
The strong APG intensifies {the magnitudes of all the terms of} $c^D_f$.
Opposition control suppresses $c^{D1}_f$ and intensifies $c^{D3}_f$ as it is designed to mitigate wall-normal convection. 

It is worth noting that~\citet{stroh_comparison_2015} {analyzed} the FIK decomposition on $c^D_{f}$ 
for on {a ZPG TBL.} 
Similarly, they reported that the terms corresponding to $c^{D1}_f$ and $c^{D3}_f$ were the dominant contributions {to the} total $c_f$. 
However, the streamwise {evolutions} of $c^{D1}_f$ and $c^{D3}_f$ in ZPG TBL were exactly opposite to what we reported.   
This difference is linked to the effect of APG, 
which intensifies the wall-normal convection{, thus changing} the balance between $c^{D1}_f$ and $c^{D3}_f$.

\begin{table}
    \begin{center}
        \def~{\hphantom{0}}
        \resizebox*{\textwidth}{!}{
        \begin{tabular}{ccccccccc} 
            Case  & $ c^T_f / c^{\rm ref}_f$  & $ c^{D1}_f / c^{\rm ref}_f$ 
            & $c^{D3}_f/ c^{\rm ref}_f$ & $ c^{P}_f/ c^{\rm ref}_f$
            & $(c^{\delta}_f + c^{D2}_f + c^{D4}_f)/ c^{\rm ref}_f$
            & $c_f / c^{\rm ref}_f$ \\ [8pt]
            
            {\rm Ref1} &$82\%$          & $210\%$           & $-126\%$     & $-84\%$      &    $19\%$ & $100\%$     \\ [3pt]
            {\rm BL1}  &$92\%$ \cred{($+10\%$)}&  $248\%$ \cred{($+38\%$)} & $-180\%$ \cgreen{($-54\%$)} & $-91\%$  \cgreen{($-7\%$)} & $18\%$ \cgreen{($-2\%$)}&  $86\%$  \cgreen{($-14\%$)}   \\ [3pt] 
            {\rm OC1}  &$71\%$ \cgreen{($-11\%$)}& $150\%$ \cgreen{($-60\%$)} & $-90\%$  \cred{($+36\%$)} & $-74\%$  \cred{($+10\%$)} & $21\%$ \cred{($+2\%$)} & $78\%$ \cgreen{($-22\%$)} \\  [6pt]

            {\rm Ref2} &$180\%$          & $1062\%$           & $-703\%$     & $-462\%$  &    $23\%$ & $100\%$      \\ [3pt]
            {\rm BL2}  &$198\%$ \cred{($+18\%$)}& $1195\%$ \cred{($+133\%$)}& $-854\%$ \cgreen{($-151\%$)}& $-488\%$ \cgreen{($-26\%$)} & $22\%$ \cgreen{($-1\%$)} & $73\%$ \cgreen{($-27\%$)}   \\ [3pt]
            {\rm OC2}  &$163\%$ \cgreen{($-17\%$)}& $1007\%$ \cgreen{($-55\%$)} & $-677\%$ \cred{($+27\%$)}  & $-431\%$ \cred{($+31\%$)} & $18\%$ \cgreen{($-5\%$)}& $81\%$ \cgreen{($-19\%$)}    \\[3pt]
           
            {\rm BD2}  &$158\%$ \cgreen{($-21\%$)}& $1018\%$ \cgreen{($-43\%$)} & $-677\%$ \cred{($+27\%$)}  & $-446\%$ \cred{($+16\%$)} & $18\%$ \cgreen{($-5\%$)}& $72\%$ \cgreen{($-28\%$)}    \\
            
       \end{tabular}
        }
       \caption{Relative contribution of the terms of the FIK decomposition to the total $c_f$ scaled by the total $c_f$ 
        for uncontrolled case ($c^{\rm ref}_f$)
        in the standard formulation and at $x/c = 0.75$. 
        The relative {changes} by the control normalized by total $c_f$ of the uncontrolled case are reported in the parentheses, 
        which is calculated as {\textit{e.g.}} $\Delta c^T_f = (c^T_f - c^{T, {\rm ref}}_f)/c^{\rm ref}_f$.
                    }
       \label{tab:FIK_075}
    \end{center}
\end{table}
Table~\ref{tab:FIK_075} summarizes the quantitative assessment of each term of the FIK decomposition at $x/c=0.75$. 
The table reports the relative contribution of the terms scaled by the total $c_f$ of the uncontrolled cases, 
such that the sum of all terms is $100\%$.

Comparison of the relative contributions between the uncontrolled cases 
indicates a significant amplification of each term's contribution in proportion, 
which is not apparent from qualitative analysis. 
This is due to the strong attenuation {of the} total $c_f$ at this location by the strong APG, 
as demonstrated in figure~\ref{fig:Cf_and_R_and_E}. 
The enhanced contributions of $c^T_f$ and $c^{P}_f$ can be confirmed as direct consequences of the intensified pressure gradient. 
Note that the terms $c^{D1}_f$ and $c^{D3}_f$ {exhibit} the most prominent changes, 
with their relative contributions increasing from $210\%$ to $1062\%$ and decreasing from $-126\%$ to $-703\%$, respectively.

At this streamwise location for NACA0012, 
the relative reduction in total $c_f$ achieved by case {\rm OC1} and {\rm BL1} is $22\%$ and $14\%$, respectively. 
However, at the same location for NACA4412, case {\rm BL2} exhibits a higher reduction of $27\%$ compared to $19\%$ for case {\rm OC2}. 
The key reason for this difference in performance lies in the response to $c^P_f$ for the control schemes. Despite changes in $c^P_f$, 
OC achieves a higher reduction in the sum of the remaining terms. 
In particular, case {\rm OC1} and {\rm OC2} achieve $32\%$ and $40\%$ reduction, respectively, 
while {\rm BL1} achieves $7\%$ and {\rm BL2} yields $1\%$. 
In a boundary layer subjected to zero-pressure gradient, 
the reduction of total $c_f$ at the same streamwise location is approximately $20\%$ 
for both OC and uniform blowing~\citep{stroh_comparison_2015, xia_direct_2015, kametani_effect_2015}.

Therefore, it can be inferred that the significant effect of {the APG} 
is responsible for the difference in the resulting reduction of $c_f$. 
However, as discussed by~\citet{atzori_uniform_2021} and~\citet{atzori_new_2023}, 
these results also raise the question of whether the modification of $c^P_f$ by applying control 
is ultimately responsible for the change in drag, 
rather than a modification of momentum transfer within the boundary layer. 
Furthermore, the high absolute {values} of each term in the uncontrolled cases {complicate} the interpretation of the standard FIK contributions, 
as the relative contributions of $c^{D1}_f$, $c^{D3}_f$, and $c^{P}_f$ are remarkably higher {when the} APG intensifies.

\subsection{Boundary-layer formulation}
{A more succinct description of the control effects in this case can be obtained with an alternative formulation of the FIK identity, which was introduced for boundary layers subjected to pressure gradients}. This formulation includes 
a contribution {containing the} mean spanwise vorticity ($c^{D\Omega}_f$) 
and a contribution {considering the} streamwise convection and pressure gradient ($c^{DP}_f$).
We refer {to the} detailed derivation outlined by~\citet{atzori_new_2023}, where 
the $c^{D\Omega}_f$ and $c^{DP}_f$ are expressed as: 
{\begin{eqnarray}
    c^{D \Omega}_f &=&  -2 \int^1_0 (1-\eta)^2 V_n \left (\frac{\partial V_n}{ \partial x_t} 
                        - \frac{\partial U_t}{ \partial y_n}  \right ) {\rm d} \eta , \\ 
    c^{D P}_f      &=&   -2 \int^1_0 (1-\eta)^2 \left ( U_t \frac{\partial U_t}{\partial x_t} 
                        + V_n \frac{\partial V_n}{\partial x_t} + \frac{\partial P}{\partial x_t} \right ) {\rm d} \eta . \\
\label{eq:new_FIK}
\end{eqnarray}}

\begin{figure}
    \centering{
        \includegraphics[width=\textwidth]{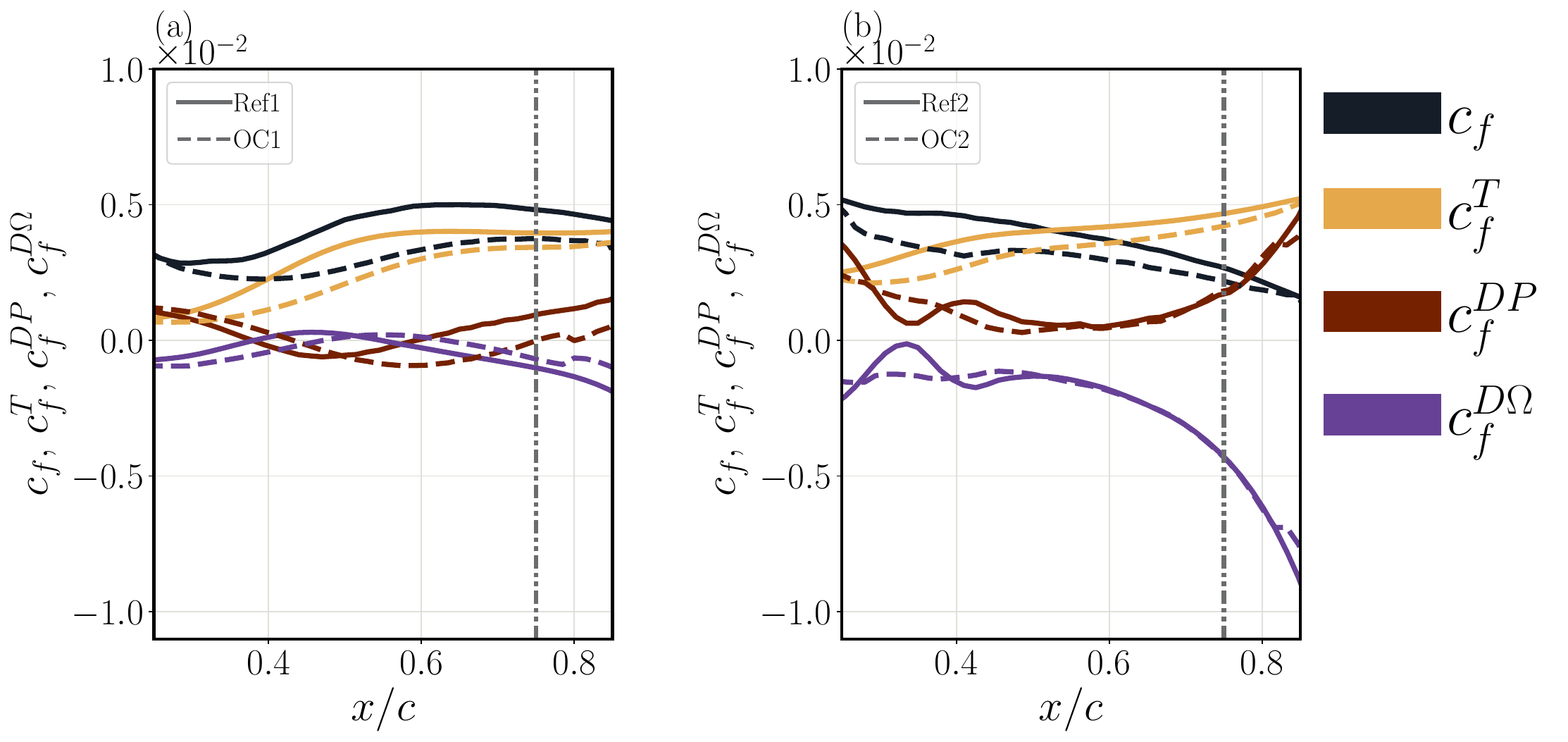}
    }
    \caption{Streamwise development of $c_f$, $c^T_f$, $c^{D \Omega}_f$ and $c^{D P}_f$ within the control section on the suction side of NACA0012 (a) and NACA4412 (b). 
    The uncontrolled and OC profiles are denoted by solid and dashed lines, respectively.
    The dash-dotted line indicates $x/c=0.75$, 
    where the relative change of components {is} reported in table~\ref{tab:FIK_NEW_075}.
    The color code of {each of the} components {is} reported at the right of the panel (b).}
    \label{fig:FIK_NEW}
\end{figure}

Note that $c^{DP}_f$ describes the balance between the evolution of the dynamic pressure in the mean flow and the pressure gradient. 
We will refer to this term as the ``dynamic-pressure contribution'' hereafter. 
The relation between the standard and boundary-layer FIK formulations is summarized as:
{\begin{equation}
    c^{D \Omega}_f + c^{D P}_f = c^{D1}_f + c^{D3}_f  + c^{P}_f,
    \label{eq:FIK_relation}
\end{equation}
}
Accordingly, $c^T_f$, $c^{D \Omega}_f$, and $c^{D P}_f$ become the dominant contributions in the FIK decomposition 
using the boundary-layer formulation. 
The streamwise evolutions of $c^T_f$, $c^{D \Omega}_f$ and $c^{D P}_f$ are illustrated in figure~\ref{fig:FIK_NEW}.
It can be observed that $c^{D P}_f$ initially decreases and then increases, 
while the absolute value of $c^{D \Omega}_f$ exhibits the same trend. 
Based on the streamwise evolution of $\beta$, the trend of $c^{DP}_f$ 
is strongly connected to the rate of change of $\beta$. 
Furthermore, the comparison between the two wing cases demonstrates 
how a stronger APG {leads to} a more rapid increase of $c^{D \Omega}_f$ and $c^{D P}_f$ in case {\rm Ref2} than in case {\rm Ref1}.

At $x/c = 0.75$, 
we report the relative componental {contributions to} $c_f$ using the boundary-layer formulation in table~\ref{tab:FIK_NEW_075}. 
\begin{table}
    \begin{center}
        \def~{\hphantom{0}}
        \begin{tabular}{cccccc} 
            Case  & $ c^T_f / c^{\rm ref}_f$  
            & $c^{DP}_f/ c^{\rm ref}_f$ & $ c^{D \Omega}_f/ c^{\rm ref}_f$ 
            & $(c^{\delta}_f + c^{D2}_f + c^{D4}_f)/ c^{\rm ref}_f$
            & $c_f / c^{\rm ref}_f$ \\ [8pt]
            
            {\rm Ref1} &$82\%$                   & $20\%$      & $-22\%$  &    $19\%$ & $100\%$     \\ [1.5pt]
            {\rm BL1}  &$92\%$ \cred{($+10\%$)}   & $33\%$  \cred{($+12\%$)} & $-56\%$  \cgreen{($-34\%$)}& $17\%$ \cgreen{($-2\%$)}&  $86\%$  \cgreen{($-14\%$)}   \\  
            {\rm OC1}  &$71\%$ \cgreen{($-11\%$)} & $1\%$ \cgreen{($-20\%$)} & $-15\%$  \cred{($+7\%$)}&  $21\%$ \cred{($+2\%$)} & $78\%$ \cgreen{($-22\%$)} \\  [6pt]

            {\rm Ref2} &$180\%$                  & $68\%$     & $-172\%$ &    $23\%$ & $100\%$      \\ [1.5pt]
            {\rm BL2}  &$198\%$ \cred{($+18\%$)}  & $108\%$ \cred{($+40\%$)}& $-257\%$ \cgreen{($-85\%$)}& $22\%$ \cgreen{($-1\%$)}& $73\%$ \cgreen{($-27\%$)}   \\ 
            {\rm OC2}  &$163\%$ \cgreen{($-17\%$)}& $71\%$ \cred{($+4\%$)} & $-173\%$ \cgreen{($-1\%$)}&  $18\%$ \cgreen{($-5\%$)}& $81\%$ \cgreen{($-19\%$)}    \\
            {\rm BD2}  &$158\%$ \cgreen{($-21\%$)}& $62\%$ \cgreen{($-6\%$)} & $-168\%$ \cred{($+4\%$)}&  $18\%$ \cgreen{($-5\%$)}& $71\%$ \cgreen{($-28\%$)}    \\
            
        \end{tabular}
        \caption{Relative contribution of the terms of the FIK decomposition to the total $c_f$ scaled by the total $c_f$ for uncontrolled case ($c^{\rm ref}_f$)
        in the boundary-layer formulation and at $x/c = 0.75$. 
        The relative {changes} by the control normalized by total $c_f$ of the uncontrolled case are reported in the parentheses, 
        which is calculated as {\textit{e.g.}} $\Delta c^T_f = (c^T_f - c^{T, {\rm ref}}_f)/c^{\rm ref}_f$.
                    }
        \label{tab:FIK_NEW_075}
    \end{center}
\end{table}
Using the {boundary-layer} formulation, 
the relative contributions exhibit lower absolute values than those obtained using the standard formulation. 
{This leads to a {clearer} identification of the most relevant control effects~\citep{atzori_new_2023}.}
For opposition control, $c^T_f$ is most significantly modified, 
while the modifications of dynamic pressure and vorticity convection are less pronounced, 
indicating that the dominant control effect of opposition control is the reduction of turbulent fluctuations. 
Similarly, body-force damping also significantly modifies $c^T_f$, but slightly decreases $c^{DP}_f$ and increases $c^{D \Omega}_f$, 
which is opposite to the effects of opposition control. 
On the other hand, uniform blowing drastically affects the dynamic pressure and vorticity convection contributions 
but has a relatively smaller impact on $c^T_f$. 
These results indicate that the modification of vorticity convection is the main mechanism affecting friction~\citep{atzori_new_2023}.

\bibliographystyle{jfm}
\bibliography{ref}

\begin{thebibliography}{55}
\expandafter\ifx\csname natexlab\endcsname\relax\def\natexlab#1{#1}\fi
\def\au#1{#1} \def\ed#1{#1} \def\yr#1{#1}\def\at#1{#1}\def\jt#1{\textit{#1}}
  \def\bt#1{#1}\def\bvol#1{\textbf{#1}} \def\vol#1{#1} \def\pg#1{#1}
  \def\publ#1{#1}\def\arxiv#1{#1}\def\org#1{#1}\def\st#1{\textit{#1}}

\bibitem[Abbas {\em et~al.\/}(2017)Abbas, Bugeda, Ferrer, Fu, Periaux,
  Pons-Prats, Valero \& Zheng]{abbas_TBLCTRLReview_2017}
{\sc \au{Abbas, Adel}, \au{Bugeda, Gabriel}, \au{Ferrer, Esteban}, \au{Fu,
  Song}, \au{Periaux, Jacques}, \au{Pons-Prats, Jordi}, \au{Valero, Eusebio} \&
  \au{Zheng, Yao}} \yr{2017}  \at{Drag reduction via turbulent boundary layer
  flow control}.  \jt{Science China Technological Sciences}  \bvol{60},
  \pg{1281--1290}.

\bibitem[Albers {\em et~al.\/}(2019)Albers, Meysonnat \&
  Schr{\"o}der]{albers_transsurfwave_2019}
{\sc \au{Albers, Marian}, \au{Meysonnat, Pascal~S} \& \au{Schr{\"o}der,
  Wolfgang}} \yr{2019}  \at{Actively reduced airfoil drag by transversal
  surface waves}.  \jt{Flow, Turbulence and Combustion}  \bvol{102},
  \pg{865--886}.

\bibitem[Anderson(1991)]{Anderson_fundamental_1991}
{\sc \au{Anderson, John~David}} \yr{1991} {\em Fundamentals of aerodynamics\/},
  2nd edn. {\em McGraw-Hill series in aeronautical and aerospace engineering\/}
  1.  \publ{New York: McGraw-Hill}.

\bibitem[Atzori {\em et~al.\/}(2023)Atzori, Mallor, Pozuelo, Fukagata, Vinuesa
  \& Schlatter]{atzori_new_2023}
{\sc \au{Atzori, Marco}, \au{Mallor, Fermín}, \au{Pozuelo, Ramón},
  \au{Fukagata, Koji}, \au{Vinuesa, Ricardo} \& \au{Schlatter, Philipp}}
  \yr{2023}  \at{A new perspective on skin-friction contributions in
  adverse-pressure-gradient turbulent boundary layers}.  \jt{International
  Journal of Heat and Fluid Flow}  \bvol{101},  \pg{109117}.

\bibitem[Atzori {\em et~al.\/}(2020)Atzori, Vinuesa, Fahland, Stroh, Gatti,
  Frohnapfel \& Schlatter]{atzori_aerodynamic_2020}
{\sc \au{Atzori, Marco}, \au{Vinuesa, Ricardo}, \au{Fahland, Georg}, \au{Stroh,
  Alexander}, \au{Gatti, Davide}, \au{Frohnapfel, Bettina} \& \au{Schlatter,
  Philipp}} \yr{2020}  \at{Aerodynamic {Effects} of {Uniform} {Blowing} and
  {Suction} on a {NACA4412} {Airfoil}}.  \jt{Flow, Turbulence and Combustion}
  \bvol{105}~(3),  \pg{735--759}.

\bibitem[Atzori {\em et~al.\/}(2021)Atzori, Vinuesa, Stroh, Gatti, Frohnapfel
  \& Schlatter]{atzori_uniform_2021}
{\sc \au{Atzori, Marco}, \au{Vinuesa, Ricardo}, \au{Stroh, Alexander},
  \au{Gatti, Davide}, \au{Frohnapfel, Bettina} \& \au{Schlatter, Philipp}}
  \yr{2021}  \at{Uniform blowing and suction applied to nonuniform
  adverse-pressure-gradient wing boundary layers}.  \jt{Physical Review Fluids}
   \bvol{6}~(11),  \pg{113904}.

\bibitem[Brunton \& Noack(2015)]{brunton_closed-loop_2015}
{\sc \au{Brunton, Steven~L.} \& \au{Noack, Bernd~R.}} \yr{2015}
  \at{Closed-{Loop} {Turbulence} {Control}: {Progress} and {Challenges}}.
  \jt{Applied Mechanics Reviews}  \bvol{67}~(5),  \pg{050801}.

\bibitem[Choi {\em et~al.\/}(1994)Choi, Moin \& Kim]{choi_active_1994}
{\sc \au{Choi, Haecheon}, \au{Moin, Parviz} \& \au{Kim, John}} \yr{1994}
  \at{Active turbulence control for drag reduction in wall-bounded flows}.
  \jt{Journal of Fluid Mechanics}  \bvol{262},  \pg{75--110}.

\bibitem[Chung \& Talha(2011)]{chung_effectivenessTCF_2011}
{\sc \au{Chung, Yongmann~M} \& \au{Talha, Tariq}} \yr{2011}  \at{Effectiveness
  of active flow control for turbulent skin friction drag reduction}.
  \jt{Physics of Fluids}  \bvol{23}~(2).

\bibitem[Dacome {\em et~al.\/}(2024)Dacome, M{\"o}rsch, Kotsonis \&
  Baars]{dacome_oppositionTBL_2024}
{\sc \au{Dacome, Giulio}, \au{M{\"o}rsch, Robin}, \au{Kotsonis, Marios} \&
  \au{Baars, Woutijn~J}} \yr{2024}  \at{Opposition flow control for reducing
  skin-friction drag of a turbulent boundary layer}.  \jt{Physical Review
  Fluids}  \bvol{9}~(6),  \pg{064602}.

\bibitem[De~Moura {\em et~al.\/}(2024)De~Moura, Machado, Dey \&
  Mukhopadhyay]{demoura_semadvantage_2024}
{\sc \au{De~Moura, BB}, \au{Machado, MR}, \au{Dey, S} \& \au{Mukhopadhyay, T}}
  \yr{2024}  \at{Manipulating flexural waves to enhance the broadband vibration
  mitigation through inducing programmed disorder on smart rainbow
  metamaterials}.  \jt{Applied Mathematical Modelling}  \bvol{125},
  \pg{650--671}.

\bibitem[Dong {\em et~al.\/}(2014)Dong, Karniadakis \&
  Chryssostomidis]{dong_robustoutlet_2014}
{\sc \au{Dong, Suchuan}, \au{Karniadakis, George~E} \& \au{Chryssostomidis,
  Chryssostomos}} \yr{2014}  \at{A robust and accurate outflow boundary
  condition for incompressible flow simulations on severely-truncated unbounded
  domains}.  \jt{Journal of Computational Physics}  \bvol{261},  \pg{83--105}.

\bibitem[Eto {\em et~al.\/}(2019)Eto, Kondo, Fukagata \&
  Tokugawa]{eto_uniformClarkY_2019}
{\sc \au{Eto, Kaoruko}, \au{Kondo, Yusuke}, \au{Fukagata, Koji} \&
  \au{Tokugawa, Naoko}} \yr{2019}  \at{Assessment of friction drag reduction on
  a clark-y airfoil by uniform blowing}.  \jt{AIAA journal}  \bvol{57}~(7),
  \pg{2774--2782}.

\bibitem[Fischer {\em et~al.\/}(2008)Fischer, Lottes \& Kerkemeier"]{nek5000}
{\sc \au{Fischer, Paul~F.}, \au{Lottes, James~W.} \& \au{Kerkemeier",
  Stefan~G.}} \yr{2008} {Nek5000} {W}eb page. {http://nek5000.mcs.anl.gov}.

\bibitem[Fukagata {\em et~al.\/}(2024)Fukagata, Iwamoto \&
  Hasegawa]{fukagata_reviewCTRL_2024}
{\sc \au{Fukagata, Koji}, \au{Iwamoto, Kaoru} \& \au{Hasegawa, Yosuke}}
  \yr{2024}  \at{Turbulent drag reduction by streamwise traveling waves of
  wall-normal forcing}.  \jt{Annual Review of Fluid Mechanics}  \bvol{56},
  \pg{69--90}.

\bibitem[Fukagata {\em et~al.\/}(2009)Fukagata, Sugiyama \&
  Kasagi]{fukagata_lower_2009}
{\sc \au{Fukagata, Koji}, \au{Sugiyama, Kazuyasu} \& \au{Kasagi, Nobuhide}}
  \yr{2009}  \at{On the lower bound of net driving power in controlled duct
  flows}.  \jt{Physica D: Nonlinear Phenomena}  \bvol{238}~(13),
  \pg{1082--1086}.

\bibitem[Ge {\em et~al.\/}(2016)Ge, Tian \& Yongqian]{ge_dynamicTCF_2017}
{\sc \au{Ge, Mingwei}, \au{Tian, De} \& \au{Yongqian, Liu}} \yr{2016}
  \at{Dynamic evolution process of turbulent channel flow after opposition
  control}.  \jt{Fluid Dynamics Research}  \bvol{49}~(1),  \pg{015505}.

\bibitem[Gad-el Hak(1996)]{gad_modern_1996}
{\sc \au{Gad-el Hak, Mohamed}} \yr{1996}  \at{{Modern Developments in Flow
  Control}}.  \jt{Applied Mechanics Reviews}  \bvol{49}~(7),  \pg{365--379}.

\bibitem[Hammond {\em et~al.\/}(1998)Hammond, Bewley \&
  Moin]{hammond_observed_1998}
{\sc \au{Hammond, E.~P.}, \au{Bewley, T.~R.} \& \au{Moin, P.}} \yr{1998}
  \at{Observed mechanisms for turbulence attenuation and enhancement in
  opposition-controlled wall-bounded flows}.  \jt{Physics of Fluids}
  \bvol{10}~(9),  \pg{2421--2423}.

\bibitem[Harun {\em et~al.\/}(2013)Harun, Monty, Mathis \&
  Marusic]{harun_PGEffect_2013}
{\sc \au{Harun, Zambri}, \au{Monty, Jason~P}, \au{Mathis, Romain} \&
  \au{Marusic, Ivan}} \yr{2013}  \at{Pressure gradient effects on the
  large-scale structure of turbulent boundary layers}.  \jt{Journal of Fluid
  Mechanics}  \bvol{715},  \pg{477--498}.

\bibitem[Hosseini {\em et~al.\/}(2016)Hosseini, Vinuesa, Schlatter, Hanifi \&
  Henningson]{hosseini_DNSwing_2016}
{\sc \au{Hosseini, Seyed~Mohammad}, \au{Vinuesa, Ricardo}, \au{Schlatter,
  Philipp}, \au{Hanifi, Ardeshir} \& \au{Henningson, Dan~S}} \yr{2016}
  \at{Direct numerical simulation of the flow around a wing section at moderate
  {Reynolds} number}.  \jt{International Journal of Heat and Fluid Flow}
  \bvol{61},  \pg{117--128}, sI\:TSFP9 special issue.

\bibitem[Hwang(1997)]{hwang_proofuniformblowing_1997}
{\sc \au{Hwang, Danny}} \yr{1997} A proof of concept experiment for reducing
  skin friction by using a micro-blowing technique.  \bt{In {\em 35th Aerospace
  Sciences Meeting and Exhibit\/}},  \pg{p. 546}.

\bibitem[Jim{\'e}nez {\em et~al.\/}(2010)Jim{\'e}nez, Hoyas, Simens \&
  Mizuno]{jimenez_TBLvsTCF_2010}
{\sc \au{Jim{\'e}nez, Javier}, \au{Hoyas, Sergio}, \au{Simens, Mark~P} \&
  \au{Mizuno, Yoshinori}} \yr{2010}  \at{Turbulent boundary layers and channels
  at moderate {Reynolds} numbers}.  \jt{Journal of Fluid Mechanics}
  \bvol{657},  \pg{335--360}.

\bibitem[Kametani {\em et~al.\/}(2015)Kametani, Fukagata, Örlü \&
  Schlatter]{kametani_effect_2015}
{\sc \au{Kametani, Yukinori}, \au{Fukagata, Koji}, \au{Örlü, Ramis} \&
  \au{Schlatter, Philipp}} \yr{2015}  \at{Effect of uniform blowing/suction in
  a turbulent boundary layer at moderate {Reynolds} number}.  \jt{International
  Journal of Heat and Fluid Flow}  \bvol{55},  \pg{132--142}.

\bibitem[King(2007)]{king_active_2007}
{\sc \au{King, Rudibert}} \yr{2007} {\em Active flow control\/}.
  \publ{Springer}.

\bibitem[King(2010)]{king_active_2010}
{\sc \au{King, Rudibert}} \yr{2010} {\em Active Flow Control II.\/}.
  \publ{Springer}.

\bibitem[Kravchenko {\em et~al.\/}(1993)Kravchenko, Choi \&
  Moin]{kravchenko_relation_1993}
{\sc \au{Kravchenko, Arthur~G}, \au{Choi, Haecheon} \& \au{Moin, Parviz}}
  \yr{1993}  \at{On the relation of near-wall streamwise vortices to wall skin
  friction in turbulent boundary layers}.  \jt{Physics of Fluids A: Fluid
  Dynamics}  \bvol{5}~(12),  \pg{3307--3309}.

\bibitem[Li {\em et~al.\/}(2021)Li, Dang, Lv \& Duan]{li_blowing-only_2021}
{\sc \au{Li, Zexiang}, \au{Dang, Xiangxin}, \au{Lv, Pengyu} \& \au{Duan,
  Huiling}} \yr{2021}  \at{Blowing-only opposition control: {Characteristics}
  of turbulent drag reduction and implementation by deep learning}.  \jt{AIP
  Advances}  \bvol{11}~(3),  \pg{035016}.

\bibitem[Maday \& Patera(1989)]{maday_spectral_1989}
{\sc \au{Maday, Yvon} \& \au{Patera, Anthony~T}} \yr{1989}  \at{Spectral
  element methods for the incompressible navier-stokes equations}.  \jt{IN:
  State-of-the-art surveys on computational mechanics (A90-47176 21-64). New
  York}  \pg{pp. 71--143}.

\bibitem[Menter(1994)]{menter_kmSST_1994}
{\sc \au{Menter, Florian~R}} \yr{1994}  \at{Two-equation eddy-viscosity
  turbulence models for engineering applications}.  \jt{AIAA journal}
  \bvol{32}~(8),  \pg{1598--1605}.

\bibitem[Monty {\em et~al.\/}(2011)Monty, Harun \&
  Marusic]{monty_APGparametric_2011}
{\sc \au{Monty, Jason~P}, \au{Harun, Zambri} \& \au{Marusic, Ivan}} \yr{2011}
  \at{A parametric study of adverse pressure gradient turbulent boundary
  layers}.  \jt{International Journal of Heat and Fluid Flow}  \bvol{32}~(3),
  \pg{575--585}.

\bibitem[Negi {\em et~al.\/}(2018)Negi, Vinuesa, Hanifi, Schlatter \&
  Henningson]{negi_unsteady_2018}
{\sc \au{Negi, Prabal~Singh}, \au{Vinuesa, Ricardo}, \au{Hanifi, Ardeshir},
  \au{Schlatter, Philipp} \& \au{Henningson, Dan~S}} \yr{2018}  \at{Unsteady
  aerodynamic effects in small-amplitude pitch oscillations of an airfoil}.
  \jt{International Journal of Heat and Fluid Flow}  \bvol{71},  \pg{378--391}.

\bibitem[Noorani {\em et~al.\/}(2016)Noorani, Vinuesa, Brandt \&
  Schlatter]{noorani_aspect_2016}
{\sc \au{Noorani, Azad}, \au{Vinuesa, Ricardo}, \au{Brandt, Luca} \&
  \au{Schlatter, Philipp}} \yr{2016}  \at{Aspect ratio effect on particle
  transport in turbulent duct flows}.  \jt{Physics of Fluids}  \bvol{28}~(11).

\bibitem[Orlandi \& Jim{\'e}nez(1994)]{orlandi_generation_1994}
{\sc \au{Orlandi, Paolo} \& \au{Jim{\'e}nez, Javier}} \yr{1994}  \at{On the
  generation of turbulent wall friction}.  \jt{Physics of Fluids}
  \bvol{6}~(2),  \pg{634--641}.

\bibitem[Pami\'es {\em et~al.\/}(2007)Pami\'es, Garnier, Merlen \&
  Sagaut]{pamies_response_2007}
{\sc \au{Pami\'es, Mathieu}, \au{Garnier, Eric}, \au{Merlen, Alain} \&
  \au{Sagaut, Pierre}} \yr{2007}  \at{Response of a spatially developing
  turbulent boundary layer to active control strategies in the framework of
  opposition control}.  \jt{Physics of Fluids}  \bvol{19}~(10),  \pg{108102}.

\bibitem[Pami\'es {\em et~al.\/}(2011)Pami\'es, Garnier, Merlen \&
  Sagaut]{pamies_opposition_2011}
{\sc \au{Pami\'es, Mathieu}, \au{Garnier, Éric}, \au{Merlen, Alain} \&
  \au{Sagaut, Pierre}} \yr{2011}  \at{Opposition control with arrayed actuators
  in the near-wall region of a spatially developing turbulent boundary layer}.
  \jt{International Journal of Heat and Fluid Flow}  \bvol{32}~(3),
  \pg{621--630}.

\bibitem[Renard \& Deck(2016)]{renard_FIKBL_2016}
{\sc \au{Renard, Nicolas} \& \au{Deck, S{\'e}bastien}} \yr{2016}  \at{A
  theoretical decomposition of mean skin friction generation into physical
  phenomena across the boundary layer}.  \jt{Journal of Fluid Mechanics}
  \bvol{790},  \pg{339--367}.

\bibitem[Schlatter \& {\"O}rl{\"u}(2010)]{schlatter_assessment_2010}
{\sc \au{Schlatter, Philipp} \& \au{{\"O}rl{\"u}, Ramis}} \yr{2010}
  \at{Assessment of direct numerical simulation data of turbulent boundary
  layers}.  \jt{Journal of Fluid Mechanics}  \bvol{659},  \pg{116--126}.

\bibitem[Skaare \& Krogstad(1994)]{skaare_turbulentTKE_1994}
{\sc \au{Skaare, Per~Egil} \& \au{Krogstad, Per-{\AA}ge}} \yr{1994}  \at{A
  turbulent equilibrium boundary layer near separation}.  \jt{Journal of Fluid
  Mechanics}  \bvol{272},  \pg{319--348}.

\bibitem[Spalart \& Watmuff(1993)]{spalart_experimental_1993}
{\sc \au{Spalart, Philippe~R} \& \au{Watmuff, Jonathan~H}} \yr{1993}
  \at{Experimental and numerical study of a turbulent boundary layer with
  pressure gradients}.  \jt{Journal of fluid Mechanics}  \bvol{249},
  \pg{337--371}.

\bibitem[Stroh {\em et~al.\/}(2012)Stroh, Frohnapfel, Hasegawa, Kasagi \&
  Tropea]{stroh_influence_2012}
{\sc \au{Stroh, Alexander}, \au{Frohnapfel, Bettina}, \au{Hasegawa, Yosuke},
  \au{Kasagi, Nobuhide} \& \au{Tropea, Cameron}} \yr{2012}  \at{The influence
  of frequency-limited and noise-contaminated sensing on reactive turbulence
  control schemes}.  \jt{Journal of Turbulence}  \bvol{13},  \pg{N16}.

\bibitem[Stroh {\em et~al.\/}(2015)Stroh, Frohnapfel, Schlatter \&
  Hasegawa]{stroh_comparison_2015}
{\sc \au{Stroh, Alexander}, \au{Frohnapfel, Bettina}, \au{Schlatter, Philipp}
  \& \au{Hasegawa, Yosuke}} \yr{2015}  \at{A comparison of opposition control
  in turbulent boundary layer and turbulent channel flow}.  \jt{Physics of
  Fluids}  \bvol{27}~(7),  \pg{075101}.

\bibitem[Stroh {\em et~al.\/}(2016)Stroh, Hasegawa, Schlatter \&
  Frohnapfel]{stroh_global_2016}
{\sc \au{Stroh, Alexander}, \au{Hasegawa, Yosuke}, \au{Schlatter, Philipp} \&
  \au{Frohnapfel, Bettina}} \yr{2016}  \at{Global effect of local skin friction
  drag reduction in spatially developing turbulent boundary layer}.
  \jt{Journal of Fluid Mechanics}  \bvol{805},  \pg{303--321}.

\bibitem[Tanarro {\em et~al.\/}(2020)Tanarro, Vinuesa \&
  Schlatter]{tanarro_effect_2020}
{\sc \au{Tanarro, \'Alvaro}, \au{Vinuesa, Ricardo} \& \au{Schlatter, Philipp}}
  \yr{2020}  \at{Effect of adverse pressure gradients on turbulent wing
  boundary layers}.  \jt{Journal of Fluid Mechanics}  \bvol{883},  \pg{A8}.

\bibitem[Vinuesa {\em et~al.\/}(2016{\natexlab{{\em a\/}}})Vinuesa, Bobke,
  {\"O}rl{\"u} \& Schlatter]{vinuesa_determining_2016}
{\sc \au{Vinuesa, Ricardo}, \au{Bobke, Alexandra}, \au{{\"O}rl{\"u}, Ramis} \&
  \au{Schlatter, Philipp}} \yr{2016{\natexlab{{\em a\/}}}}  \at{On determining
  characteristic length scales in pressure-gradient turbulent boundary layers}.
   \jt{Physics of fluids}  \bvol{28}~(5).

\bibitem[Vinuesa {\em et~al.\/}(2017)Vinuesa, Hosseini, Hanifi, Henningson \&
  Schlatter]{vinuesa_wingTBL_2017}
{\sc \au{Vinuesa, Ricardo}, \au{Hosseini, Seyed~M}, \au{Hanifi, Ardeshir},
  \au{Henningson, Dan~S} \& \au{Schlatter, Philipp}} \yr{2017}
  \at{Pressure-gradient turbulent boundary layers developing around a wing
  section}.  \jt{Flow, turbulence and combustion}  \bvol{99},  \pg{613--641}.

\bibitem[Vinuesa {\em et~al.\/}(2022)Vinuesa, Lehmkuhl, Lozano-Dur{\'a}n \&
  Rabault]{vinuesa_flow_2022}
{\sc \au{Vinuesa, Ricardo}, \au{Lehmkuhl, Oriol}, \au{Lozano-Dur{\'a}n, Adrian}
  \& \au{Rabault, Jean}} \yr{2022}  \at{Flow control in wings and discovery of
  novel approaches via deep reinforcement learning}.  \jt{Fluids}
  \bvol{7}~(2),  \pg{62}.

\bibitem[Vinuesa {\em et~al.\/}(2018)Vinuesa, Negi, Atzori, Hanifi, Henningson
  \& Schlatter]{vinuesa_turbulent_2018}
{\sc \au{Vinuesa, Ricardo}, \au{Negi, Prabal~Singh}, \au{Atzori, Marco},
  \au{Hanifi, Ardeshir}, \au{Henningson, Dan~S} \& \au{Schlatter, Philipp}}
  \yr{2018}  \at{Turbulent boundary layers around wing sections up to {R} e c =
  1 , 000 , 000}.  \jt{International Journal of Heat and Fluid Flow}
  \bvol{72},  \pg{86--99}.

\bibitem[Vinuesa {\em et~al.\/}(2016{\natexlab{{\em b\/}}})Vinuesa, Prus,
  Schlatter \& Nagib]{vinuesa_ductconvergence_2016}
{\sc \au{Vinuesa, Ricardo}, \au{Prus, Cezary}, \au{Schlatter, Philipp} \&
  \au{Nagib, Hassan~M}} \yr{2016{\natexlab{{\em b\/}}}}  \at{Convergence of
  numerical simulations of turbulent wall-bounded flows and mean cross-flow
  structure of rectangular ducts}.  \jt{Meccanica}  \bvol{51},
  \pg{3025--3042}.

\bibitem[Vinuesa \& Schlatter(2017)]{vinuesa_skin_2017}
{\sc \au{Vinuesa, Ricardo} \& \au{Schlatter, Philipp}} \yr{2017} Skin-friction
  control of the flow around a wing section through uniform blowing.  \bt{In
  {\em Proceedings of European Drag Reduction and Flow Control Meeting
  (EDRFCM)\/}}.

\bibitem[Viswanath(2002)]{viswanath_riblets_2002}
{\sc \au{Viswanath, PR}} \yr{2002}  \at{Aircraft viscous drag reduction using
  riblets}.  \jt{Progress in Aerospace Sciences}  \bvol{38}~(6-7),
  \pg{571--600}.

\bibitem[Wang {\em et~al.\/}(2024)Wang, Animasaun, Muhammad \&
  Okoya]{wang_reviewCTRL_2024}
{\sc \au{Wang, Fu~Zhang}, \au{Animasaun, IL}, \au{Muhammad, Taseer} \&
  \au{Okoya, SS}} \yr{2024}  \at{Recent advancements in fluid dynamics: drag
  reduction, lift generation, computational fluid dynamics, turbulence
  modelling, and multiphase flow}.  \jt{Arabian Journal for Science and
  Engineering}  \pg{pp. 1--13}.

\bibitem[Wang {\em et~al.\/}(2016)Wang, Huang \& Xu]{wang_active_2016}
{\sc \au{Wang, Yin-Shan}, \au{Huang, Wei-Xi} \& \au{Xu, Chun-Xiao}} \yr{2016}
  \at{Active control for drag reduction in turbulent channel flow: the
  opposition control schemes revisited}.  \jt{Fluid Dynamics Research}
  \bvol{48}~(5),  \pg{055501}.

\bibitem[Xia {\em et~al.\/}(2015)Xia, Huang, Xu \& Cui]{xia_direct_2015}
{\sc \au{Xia, Qian-Jin}, \au{Huang, Wei-Xi}, \au{Xu, Chun-Xiao} \& \au{Cui,
  Gui-Xiang}} \yr{2015}  \at{Direct numerical simulation of spatially
  developing turbulent boundary layers with opposition control}.  \jt{Fluid
  Dynamics Research}  \bvol{47}~(2),  \pg{025503}.

\bibitem[Yousif {\em et~al.\/}(2023)Yousif, Yu, Hoyas, Vinuesa \&
  Lim]{yousif_ML_GAN_2023}
{\sc \au{Yousif, Mustafa~Z}, \au{Yu, Linqi}, \au{Hoyas, Sergio}, \au{Vinuesa,
  Ricardo} \& \au{Lim, HeeChang}} \yr{2023}  \at{A deep-learning approach for
  reconstructing 3d turbulent flows from 2d observation data}.  \jt{Scientific
  Reports}  \bvol{13}~(1),  \pg{2529}.

\end{thebibliography}
\end{document}